\documentclass[aps,prd,preprint,groupedaddress,11pt,preprintnumbers,nofootinbib]{revtex4-1}

\usepackage{amsmath}
\usepackage{amssymb}
\usepackage{graphicx}
\usepackage{subfig}
\usepackage{braket}
\usepackage{slashed}
\usepackage{paralist}

\newcommand{\ra}{\rightarrow}
\newcommand{\lra}{\leftrightarrow}

\begin{document}

\preprint{TUM-HEP-890/13}

\title{$D$-meson lifetimes within the heavy quark expansion}


\author{Alexander Lenz}
\email[]{alexander.lenz@cern.ch}
\affiliation{IPPP, Department of Physics, University of Durham, DH1 3LE, UK}
\affiliation{CERN - Theory Divison, PH-TH, Case C01600, CH-1211 Geneva 23, Switzerland}

\author{Thomas Rauh}
\email[]{thomas.rauh@mytum.de}
\affiliation{TU M\"unchen, Physik-Department, 85748 Garching, Germany}


\date{\today}

\begin{abstract}
Even if new data indicate that direct CP violation in $D$-meson decays is compatible with the standard model expectation, the first hints for direct
CP violation have triggered a lot of interest, and charm phenomenology will remain an essential part of new physics searches due to its unique role
as a probe for flavor-changing neutral currents among up-type quarks.
Charm physics poses considerable theoretical challenges, because the charm mass is neither light nor truly heavy. The heavy quark expansion
(HQE) provides a perturbative expansion in the inverse heavy quark mass for inclusive rates. It has proved to be very successful in the $B$ sector,
yet its validity for charm decays has often been questioned.
We present results of a HQE study of $D$-meson lifetimes including NLO QCD and subleading \(1/m_c\) corrections. We find good agreement with
experimental data, but with huge hadronic uncertainties due to missing lattice input for hadronic matrix elements. 
\end{abstract}


\maketitle

\section{Introduction\label{sec:introduction}}
The charm quark plays a unique role in the standard model. Since the top quark decays before it can hadronize \cite{bigi86},
charm is the only up-type quark, whose hadronic weak decays can be analyzed. The $D$ sector thus offers the only handle to probe flavor-changing
neutral currents among weak-isospin up quarks. Mixing is by now well established in the charm sector
\cite{bediaga12,lhcb12,hfag12} and has already provided severe constraints on some new physics models \cite{isidori10}.
First experimental results on CP violation in \(D^0\ra\pi^+\pi^-,K^+K^-\) decays \cite{lhcb11} caused a lot of attention among phenomenologists \cite{bigi12};
see e.g. Ref. \cite{lhcb12} for an overview.
However, after a recent update \cite{lhcb13}, the experimental results seem to be compatible with the standard model expectation. Yet, the present experimental average
\cite{hfag12} for \(\Delta a_{\text{CP}}^{\text{dir}}\) still differs from zero by \(2.7\sigma\), and further analyses are mandatory to resolve this issue.
Unfortunately, there are severe theoretical challenges in the charm sector, because the charm quark mass is neither light nor truly heavy.\\
We present a study of $D$-meson lifetimes within the heavy quark expansion (HQE) \cite{HQE1,HQE2,HQE3,HQE4,HQE5,HQE6,HQE7,HQE8,HQE9},
an operator product expansion (OPE)-based framework \cite{wilson69}, that expresses inclusive decay rates as an expansion in the inverse heavy quark mass.
Lifetimes are used for the purpose of probing the HQE in charm, because new physics effects are expected to be negligible.
This formalism is well established and experimentally verified in the $B$ sector. The validity of the HQE in the $D$ sector has, however, often been questioned,
because of the lower charm quark mass. But there is a simple yet persuasive argument that suggests that the situation is not that pessimistic \cite{bobrowski12}.
The HQE is an expansion in the hadronic scale $\Lambda$ over the momentum release $\sqrt{m_{\text{i}}^2-m_{\text{f}}^2}$ in the considered decay rate
$\text{i}\ra\text{f}$. The confrontation of the HQE prediction for the lifetime difference in the neutral $B_s$ meson system, $\Delta \Gamma_s$ \cite{lenz11},
\footnote{This is based on the computations in Refs. \cite{beneke96,beneke98,beneke02,beneke03,lenz06}.}
with recent experimental results \cite{hfag12} shows excellent agreement:
\begin{equation}
 \Delta\Gamma_s^{\text{SM}}=(0.087\pm0.021)\ \mbox{ps}^{-1},\hspace{1.5cm}\Delta\Gamma_s^{\text{exp}}=(0.081\pm0.011)\ \mbox{ps}^{-1}.
\end{equation}
The dominant contribution to \(\Delta\Gamma_s\) comes from the \(D_s^{(*)+}D_s^{(*)-}\) final state, where the momentum release is \(\sim 3.3-3.6\text{ GeV}\).
Explicit calculation shows that the HQE expansion parameter for \(\Delta\Gamma_s\) is around 1/5 \cite{lenz12}. This implies that the relevant hadronic scale is
of order $0.7\text{ GeV}$ and thus slightly below the \(1\text{ GeV}\) it is commonly expected to be. Comparison with the typical momentum release
in $D$-meson decays should yield a rough estimate of the expansion parameter governing the HQE in the $D$ sector. For \(D^0\) and \(D^+\) mesons,
the dominant final states consist of a kaon and one to three pions, which corresponds to a momentum release of \(\sim1.6-1.8\text{ GeV}\).
For \(D_s^+\), the dominant decay channels are a kaon pair and one or two pions, as well as \(\eta'(958)\pi^+\) and \(\eta\rho^+\) with a momentum release of
\(\sim 1.5-1.6\text{ GeV}\), but there is also a large branching ratio to \(\eta'(958)\rho^+\) with a momentum release of just \(\sim0.9\text{ GeV}\) \cite{pdg}.
This suggests an expansion parameter of $\sim0.4-0.5$ which looks rather promising. Yet it is possible that final states with small momentum release like
$\eta'(958)\rho^+$ spoil the validity of the HQE in the case of $D_s^+$.\\
A calculation of subleading corrections in charm mixing within the framework of the HQE \cite{bobrowski11} likewise did not show signs of a breakdown
of the perturbative approach. It turned out that the charm width difference receives NLO QCD corrections at a level below \(50\%\) and \(1/m_c\)
corrections of \(30\%\). Thus, we consider it worthwhile to investigate $D$-meson lifetimes within the framework of the HQE.\\
The outline of this paper is as follows. In Sec. \ref{sec:history} we summarize previous work on $D$-meson lifetimes. The relevant formulas for
the HQE are given in Sec. \ref{sec:HQE}. In Sec. \ref{sec:phenomenology} we present a phenomenological analysis of the lifetimes of charmed mesons.
We conclude in Sec. \ref{sec:conclusion}.
%
%
%
%
\section{History of $D$-meson Lifetimes\label{sec:history}}
\begin{figure}[htbc]
\centering
\subfloat[]{
        \includegraphics[width=0.28\textwidth]{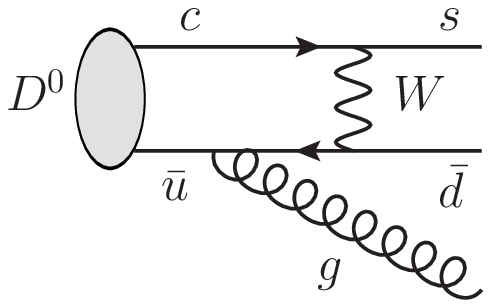}
        \label{fig:WAgluon1}
}
\quad
\subfloat[]{
        \includegraphics[width=0.28\textwidth]{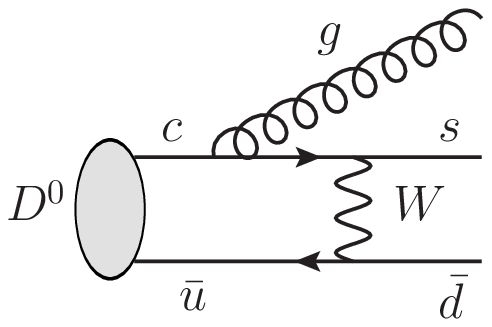}
        \label{fig:WAgluon2}
    }
\quad
\subfloat[]{
        \includegraphics[width=0.28\textwidth]{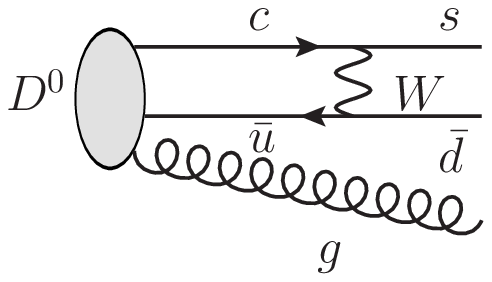}
        \label{fig:WAgluon3}
    }
\caption{Gluon emission from the weak annihilation diagram.\label{fig:WAgluon}}
\end{figure}

\begin{figure}[t]
\centering
  \includegraphics[width=0.4\textwidth]{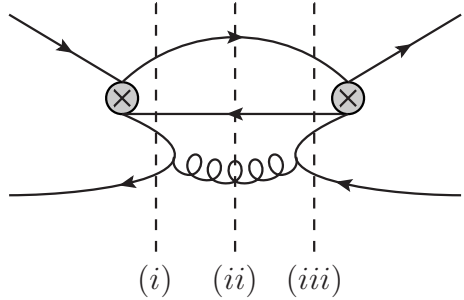}
\caption{Different cuts contributing to the weak annihilation. The \(f_D^2/\Braket{E_{\bar{u}}}^2\)-enhanced term due to the cut (ii) considered in Refs.
\cite{bander80,fritzsch80,bernreuther80} is canceled by interference effects (i) and (iii), such that the fully inclusive rate experiences the
correct \(1/m_c^3\) scaling behavior predicted by the HQE.}
\label{fig:WAcuts}
\end{figure}

The first estimations of the lifetimes of charmed particles were based on the assumption that the free charm decay dominates the process, while the
lighter quarks in the hadron only act as spectators \cite{gaillard75,ellis75,cabibbo78}.
Within this spectator picture, the lifetimes of all charmed mesons are expected to be nearly identical. Thus, it
came as quite a surprise when the first data showed that the lifetimes substantially differed, especially since the first
measurements hinted at a much larger deviation than what is established today \cite{bacino80,allasia80,ushida80}.
As a response, two mechanisms were suggested trying to explain this effect. The first proposed a reduction of the \(D^+\) decay rate due to the
Pauli interference contribution shown in Fig. \ref{fig:PI} \cite{guberina79}. Here, the \(1/m_c^3\) suppression of the Pauli interference
effect was not accounted for, i.e. in today's language the authors have set \(16\pi^2\left(f_D^2M_D\middle/m_c^3\right)=1\).
The second mechanism proposed an enhancement of the \(D^0\) decay rate due to the weak annihilation diagram in Fig. \ref{fig:WA}.
The weak annihilation contribution suffers from chirality suppression. To ease this suppression, it was proposed in Refs.
\cite{bander80,fritzsch80,bernreuther80} to consider gluon emission from the ingoing quark lines as illustrated in Fig. \ref{fig:WAgluon}.
This yields a contribution proportional to \(f_D^2/\Braket{E_{\bar{q}}}^2\), where \(f_D\approx200\text{ MeV}\) is the $D$-meson decay
constant and \(\Braket{E_{\bar{q}}}\) denotes the average energy of the initial antiquark.
In Ref. \cite{bernreuther80}, the authors additionally included the Cabibbo-suppressed weak annihilation of \(D^+\) and obtained for the effects
of weak annihilation in \(D^0\) and \(D^+\)
\begin{equation}
 \left.\frac{\tau(D^+)}{\tau(D^0)}\right|_{\text{\cite{bernreuther80}}}\approx5.6-6.9, \hspace{1cm}\left.\frac{\tau(D^+)}{\tau(D^0)}\right|_{\text{PDG'12\cite{pdg}}}=2.536\pm0.019.
\end{equation}
One should keep in mind that Pauli interference, which is now known to be the dominant effect, is still neglected here. This shows what a severe
overestimation these early analyses were.\\
Further studies of the Pauli interference effect \cite{kobayashi81} already obtained results similar to the later
HQE treatments; however, they were still in a less formal fashion. The first systematic treatments were performed in the following years,
when the idea of HQE was developed and was applied to charm decays \cite{HQE1,bilic84,HQE2}. The formula below represents
the starting point of the HQE and was first presented in Ref.  \cite{HQE1} with a sign error which was corrected in Ref. \cite{bilic84}
\begin{equation}
\begin{aligned}
 \Gamma(D^+)=\frac{G_F^2}{2M_D}\Bra{D^+}&\frac{m_c^5}{64\pi^3}\frac{2C_+^2+C_-^2}{3}\bar{c}c+\frac{m_c^2}{2\pi}
 \Big[\left(C_+^2+C_-^2\right)(\bar{c}\Gamma_\mu T^A d)(\bar{d}\Gamma_\mu T^A c)\\
 & +\frac{2C_+^2-C_-^2}{3}(\bar{c}\Gamma_\mu d)(\bar{d}\Gamma_\mu c)\Big]\Ket{D^+}.
 \end{aligned}
\end{equation}
We have rewritten this in the color-singlet and color-octet basis commonly used today for \(\Delta C=0\) operators.
The leading term describes the decay of the free charm quark in the parton model, and the following term describes the \(1/m_c^3\)-suppressed effect of
Pauli interference. Neglecting weak annihilation, the total decay rate for \(D^0\) is given by the first term of this expression.
The four quark operators have been evaluated in the vacuum insertion approximation. In the early analyses, the lifetime ratios were generally underestimated,
\begin{equation}
 \left.\frac{\tau(D^+)}{\tau(D^0)}\right|_\text{early HQE analyses}\approx1.5,
\end{equation}
which was mainly due to a too-small estimate for the decay constant \(f_D\approx160-170\text{ MeV}\). The present value of \(f_D=212.7\text{ MeV}\)
yields \(\left.\tau(D^+)\middle/\tau(D^0)\right.\approx2.2\), which drastically improves the consistency with experiments. In Ref. \cite{shifman86}, the effects
of hybrid renormalization were first included. This constitutes the present state of theory predictions for the ratio of \(D^+\) and \(D^0\) lifetimes.
It was argued \cite{HQE2,shifman86} that \(\tau(D_s^+)\approx\tau(D^0)\), which contradicted the experimental situation at that time.
However, better experimental results quickly straightened out the charmed mesons' lifetimes.
It was further shown in Ref. \cite{guberina86} that the HQE was able to correctly reproduce the hierarchy of lifetimes in the charm sector:
\begin{equation}
 \tau(D^+)>\tau(D^0)>\tau(\Xi_c^+)>\tau(\Lambda_c^+)>\tau(\Xi_c^0)>\tau(\Omega_c^0).
\end{equation}
During the second half of the 1980s, the experimental values improved and got a lot closer to the present data. In 1992,
Bigi and Uraltsev explained the apparent contradiction \cite{HQE5} between the \(1/m_c\) scaling of the HQE
and the \(f_D^2/\Braket{E_{\bar{q}}}^2\)-enhanced gluon bremsstrahlung of Refs. \cite{bander80,fritzsch80,bernreuther80}. They showed that
these power-enhanced terms cancel in fully inclusive rates between different cuts as indicated in Fig. \ref{fig:WAcuts} and preasymptotic
effects hence scale with \(1/m_c^3\), consistently with the HQE. This was later confirmed by an explicit calculation, first for \(\Delta\Gamma_s\)
in Ref. \cite{beneke98} and then for lifetimes in Refs. \cite{beneke02,ROME}.
In the following, Bigi and Uraltsev applied the HQE to charm lifetimes \cite{bigi93a,bigi94}. For the \(D_s^+\) meson they found
\begin{equation}
  \left.\frac{\tau(D_s^+)}{\tau(D^0)}\right|_{\text{\cite{bigi93a}}}=0.9-1.3, \hspace{1cm}\left.\frac{\tau(D_s^+)}{\tau(D^0)}\right|_{\text{PDG'12\cite{pdg}}}=1.219\pm0.018,
\end{equation}
where the uncertainty dominantly arises from the weak annihilation. However, during the establishment of \(1/m_Q\) expansions,
the theory focus shifted towards B physics, where the corrections are smaller and better controlled \cite{bigi94a} (see Ref. \cite{lenz11} for updated NLO results).
The validity of the HQE in charm decays has frequently been questioned since, because of the smaller charm quark
mass. Yet, it has been shown in a number of reviews by Bigi \textit{et} al. that the lifetimes of weakly decaying charmed hadrons can
be accounted for within the HQE at least in a "semiquantitative" fashion \cite{bigi95,bigi97,bianco03}.\\
\\
Summing up, the HQE was successful in reproducing the observed pattern of charm hadron lifetimes and explaining the
issue of gluonic bremsstrahlung enhancement.
However, charm lifetimes have so far only been considered at leading order in QCD. Subleading \(1/m_c\) corrections to the spectator
effects were never studied in charm, although they are expected to be sizeable. There has never been a dedicated quantitative analysis
of \(\tau(D^+)/\tau(D^0)\). For the numerical estimations, the vacuum saturation approximation of the four quark operators has been invoked. 
Deviations from this were parametrized in Refs. \cite{bigi94,bigi94a}, but never quantitatively examined in the charm sector.
Also, the mass of the strange quark and the muon have generally been neglected. We aim to improve on this in a number of crucial points:
\begin{itemize}
 \item We include NLO QCD corrections, which considerably reduces the dependence on the renormalization scale. This required a NLO
 computation of the coefficients for the semileptonic weak annihilation in \(D_s^+\) presented in Appendix \ref{sec:SLWA}.
 \item Bag parameters are introduced to allow for the matrix elements to differ from their vacuum insertion approximation value.
 \item We compute subleading \(1/m_c^4\) corrections to the spectator effects to investigate the convergence behavior of
 the HQE.
 \item The effects of the strange quark and muon mass are fully included in the phase-space factors.
\end{itemize}
This improves the theory predictions for the lifetimes of $D$ mesons considerably.
%
%
%
%
%
\section{Inclusive rates for charmed hadrons\label{sec:HQE}}
\begin{figure}[htbc]
\centering
\subfloat[Pauli Interference]{
        \includegraphics[width=0.28\textwidth]{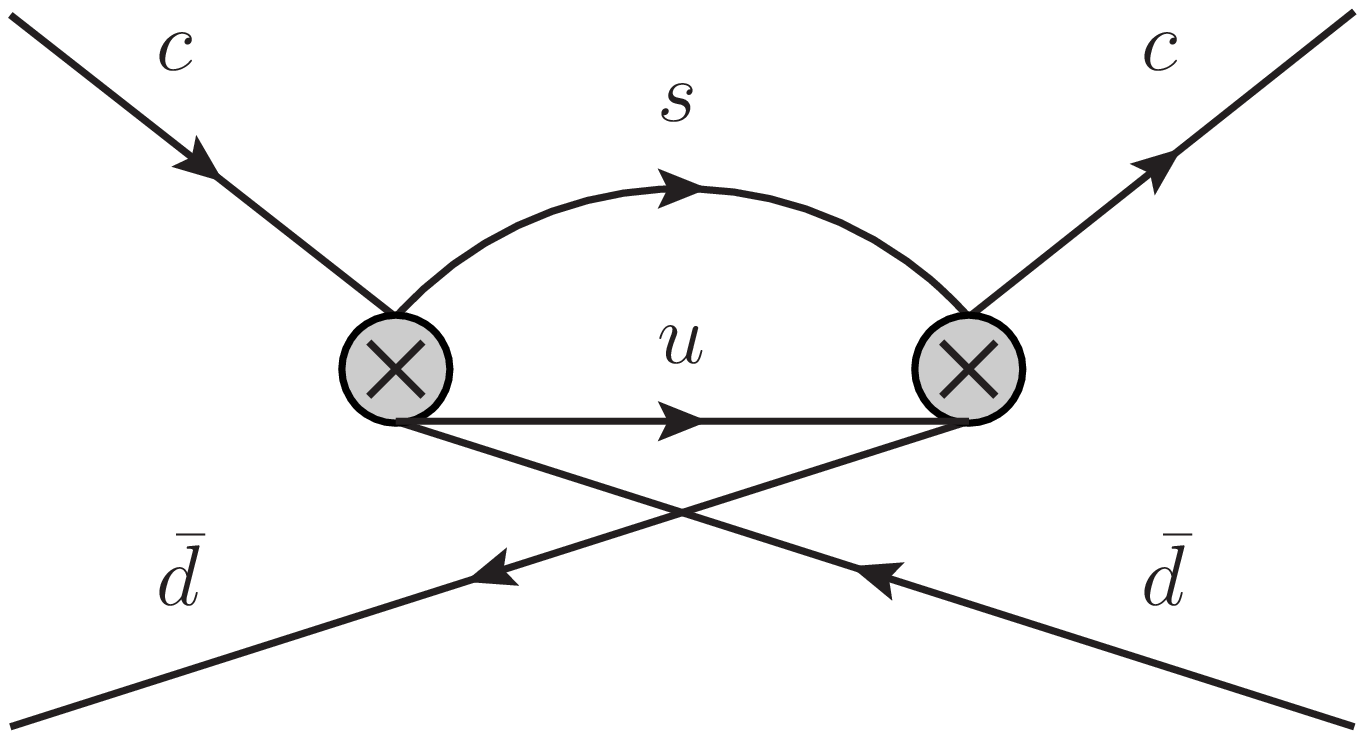}
        \label{fig:PI}
}
\quad
\subfloat[Weak Annihilation in \(D^0\)]{
        \includegraphics[width=0.28\textwidth]{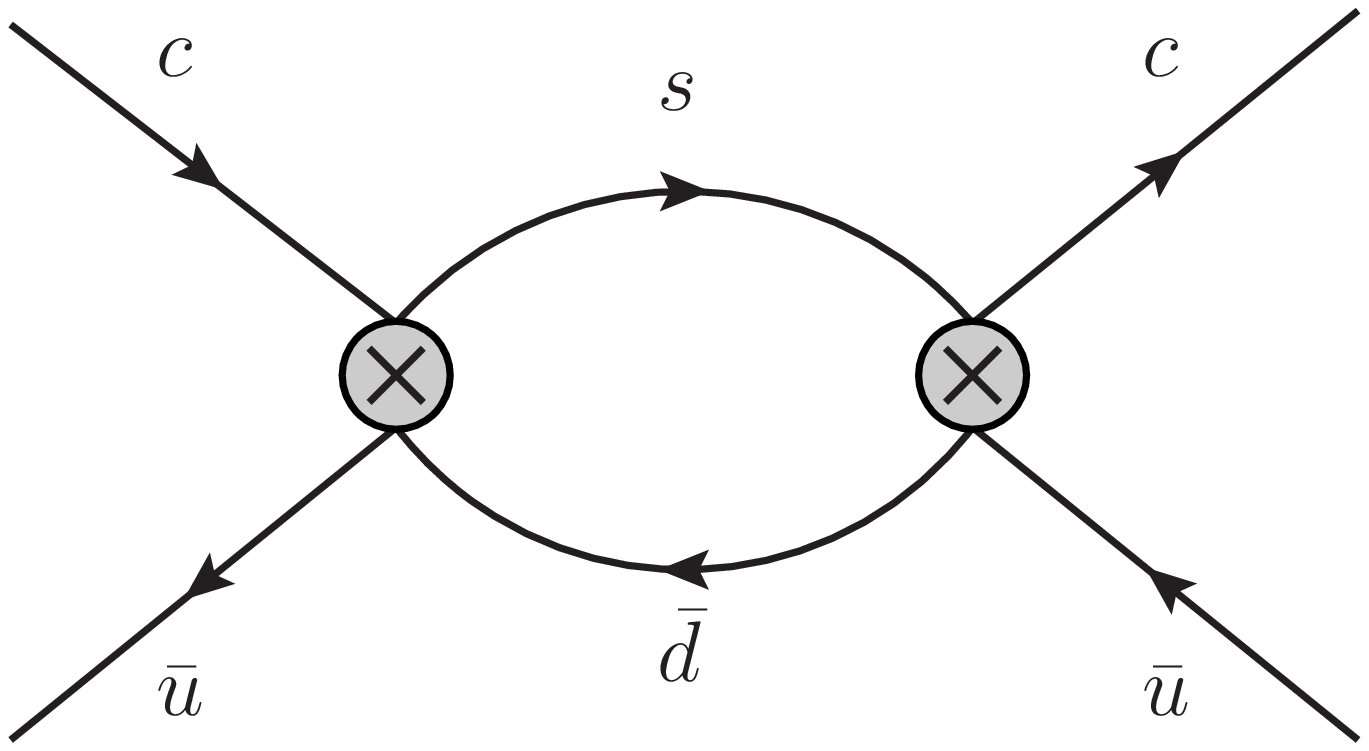}
        \label{fig:WA}
    }
\quad
\subfloat[Weak Annihilation in \(D_s^+\)]{
        \includegraphics[width=0.28\textwidth]{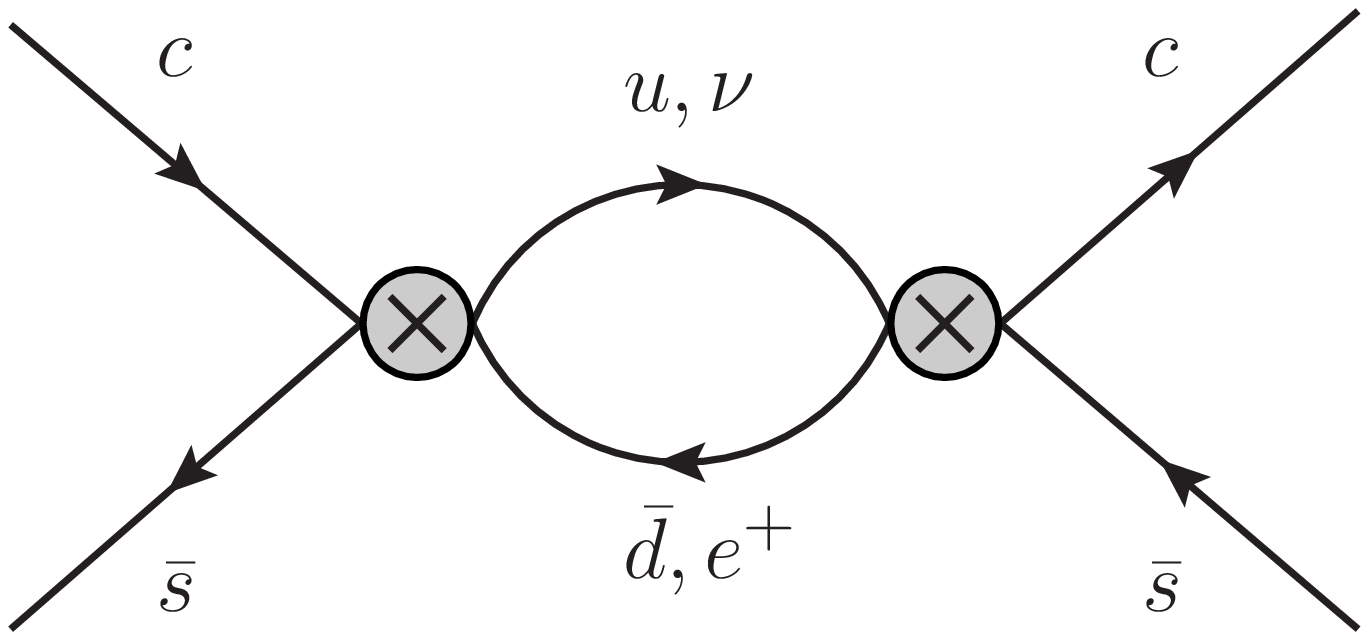}
        \label{fig:WAp}
    }
\caption{Spectator contributions to the lifetimes of charmed hadrons.\label{fig:Gamma3}}
\end{figure}

\begin{figure}[htbc]
\centering
   \includegraphics[width=0.3\textwidth]{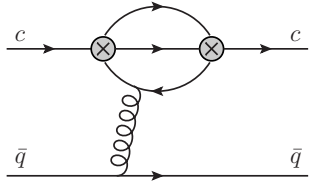}
\caption{Sample diagram for \(\mathcal{T}_3^{\text{sing}}\).\label{fig:Tsing}}
\end{figure}

\begin{figure}[t]
\centering
\subfloat[Standard contraction]{
        \includegraphics[width=0.4\textwidth]{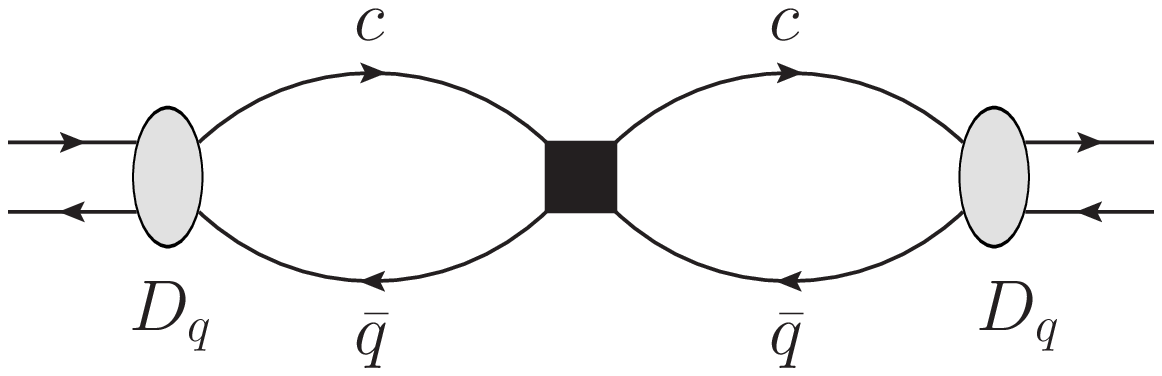}
        \label{fig:standardcontraction}
}
\quad
\subfloat[Penguin/Eye contraction]{
        \includegraphics[width=0.4\textwidth]{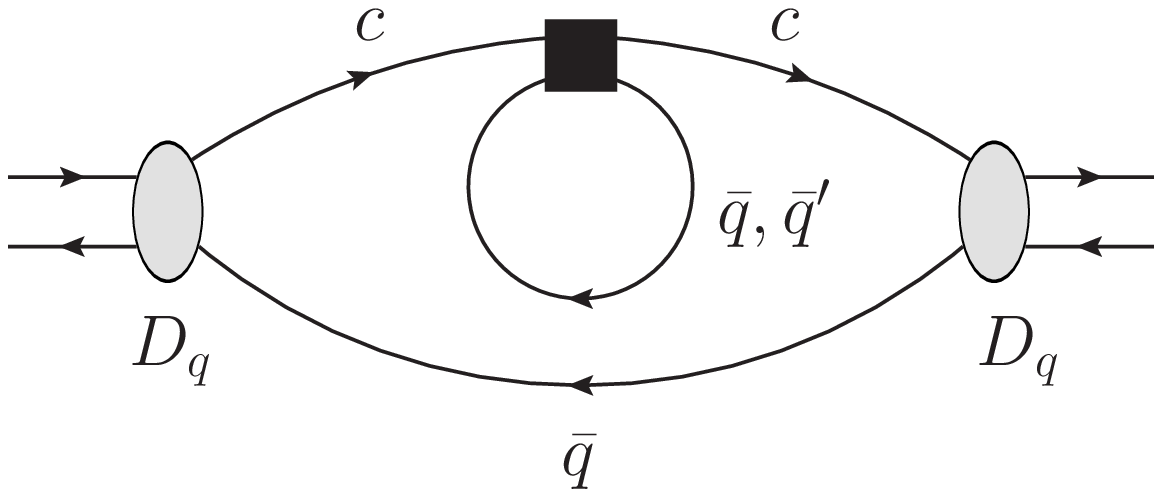}
        \label{fig:eyecontraction}
    }
\caption{Relevant Feynman diagrams for the nonperturbative matrix elements of $D$ mesons.}
\label{fig:contractions}
\end{figure}

The HQE provides an OPE-based framework for the description of inclusive decay rates of hadrons containing one heavy quark
\cite{HQE1,HQE2,HQE3,HQE4,HQE5,HQE6,HQE7,HQE8,HQE9}.
It yields an expansion of \(\Gamma(H_Q)\) in \(\left.\Lambda\middle/m_Q\right.\), where \(\Lambda\) denotes the hadronic scale expected to be of order \(\Lambda_\text{QCD}\)
and \(m_Q\) denotes the heavy quark mass. The HQE is based on the concept of quark hadron duality \cite{poggio76}. We work under the assumption that duality holds and
then confront the phenomenological results with experimental data.\\
Integrating out the $W$ boson, one obtains the following effective Hamiltonian describing \(\Delta C=1\) transitions (see e.g. Ref. \cite{buchalla95} for a review):
\begin{equation}
 \mathcal{H}_{\text{eff}}=\frac{G_F}{\sqrt{2}}\left[C_1(\mu_1)Q_1+C_2(\mu_1)Q_2+Q_e+Q_\mu\right]+\text{h.c.}.
\end{equation}
The local \(\Delta C=1\) operators are
\begin{equation}
 Q_1=\bar{s}_j'\gamma_\mu(1-\gamma_5)c_i\,\bar{u}_i\gamma^\mu(1-\gamma_5)d_j',\hspace{1cm} Q_2=\bar{s}_i'\gamma_\mu(1-\gamma_5)c_i\,\bar{u}_j\gamma^\mu(1-\gamma_5)d_j'
 \label{eq:DeltaC1ops}
\end{equation}
and
\begin{equation}
 Q_{l}=\bar{s}'\gamma_\mu(1-\gamma_5)c\,\bar{l}\gamma^\mu(1-\gamma_5)\nu,
\end{equation}
with \(d'=V_{ud}d+V_{us}s\) and \(s'=V_{cs}s+V_{cd}d\). The Wilson coefficients \(C_i\) have been computed at NLO QCD in Refs. \cite{altarelli81,buras90}
and at NNLO QCD in Ref. \cite{gorbahn04}. We will, however, only use the NLO expressions in the NDR scheme defined in Ref. \cite{buras90} throughout this work.
The HQE then integrates out the hard momenta of the final-state particles.
We use the optical theorem to express the decay rate via the imaginary part of the forward scattering amplitude
\begin{equation}
 \Gamma(H_c)=\frac{1}{2M_{H_c}}\Braket{H_c|\Im\left(i\int d^4xT\left[\mathcal{H}_{\text{eff}}(x)\mathcal{H}_{\text{eff}}(0)\right]\right)|H_c}
 =\frac{1}{2M_{H_c}}\Braket{H_c|\mathcal{T}|H_c}.
\label{eq:HQE}
\end{equation}
For small \(x\), i.e. large energy release, the transition operator \(\mathcal{T}\) can then be expanded by an OPE \cite{wilson69}. The result is a series
\begin{equation}
 \mathcal{T}=\mathcal{T}_0+\mathcal{T}_2+\mathcal{T}_3+\mathcal{T}_4+\dots,
\end{equation}
where \(\mathcal{T}_n\) denotes the \(1/m_c^n\) suppressed part of \(\mathcal{T}\). The leading term \(\mathcal{T}_0=\sum c_3^{(f)}\bar{c}c\) describes the free charm decay.
Here, no nonperturbative contributions are present, since the hadronic matrix element \(\Braket{H_c|\bar{c}c|H_c}=1+\mathcal{O}\left(1\middle/m_c^2\right)\) is trivial.
The corresponding Wilson coefficients \(c_3^{(f)}\) have been determined for b decays in Ref. \cite{bdecays} and can be easily adjusted to the charm sector.
To this order, the lifetimes of all weakly decaying charmed hadrons are equal. We observe that no \(\mathcal{T}_1\) term is present, because the contribution of the respective
operator \(\bar{c}i\slashed{D}c\) can be incorporated in the leading term \(\mathcal{T}_0\) by application of the equations of motion.
The \(1/m_c^2\)-suppressed part takes the form \cite{HQE6}
\begin{equation}
 \frac{1}{2M_{H_c}}\Braket{H_c|\mathcal{T}_2|H_c}=\sum c_3^{(f)}\frac{\mu_G^2(H_c)-\mu_\pi^2(H_c)}{2m_c^2}+\sum 2c_5^{(f)}\frac{\mu_G^2(H_c)}{m_c^2},
\end{equation}
where the first term originates from the heavy quark effective theory (HQET) expansion of the dimension-3 matrix element \(\Braket{H_c|\bar{c}c|H_c}\). The hadronic parameters \(\mu_\pi^2(H_c)\) and
\(\mu_G^2(H_c)\) are the matrix elements of the kinetic and chromomagnetic operators, respectively. To this order, lifetime differences between charmed mesons only arise through
\(SU(3)\) flavor breaking of the hadronic matrix elements. The dominant contributions originate at order \(1/m_c^3\) from the Pauli interference and weak annihilation diagrams 
shown in Fig. \ref{fig:Gamma3}.
They describe \(2\ra2\) instead of \(1\ra3\) processes, and hence are phase-space enhanced by a factor of \(16\pi^2\). We neglect further contributions of order \(1/m_c^3\) that lack
this enhancement. We decompose the \(\left.1\middle/m_c\right.^3\) part of the transition operator as
\begin{equation}
 \mathcal{T}_3=\mathcal{T}_3^{\text{PI}}+\mathcal{T}_3^{\text{WA}_0}+\mathcal{T}_3^{\text{WA}_+}+\mathcal{T}_3^{\text{sing}}.
\end{equation}
The contribution \(\mathcal{T}_3^{\text{sing}}\) arises from strong interactions of the free quark decay with the spectator quark of the type shown in Fig. \ref{fig:Tsing}.
The effect of \(\mathcal{T}_3^{\text{sing}}\) cancels in the considered lifetime ratios in the limit of \(SU(3)\) flavor symmetry since the corresponding
dimension-6 operators are \(SU(3)\) flavor singlets.
The remaining terms are
\begin{equation}
\begin{aligned}
&\mathcal{T}_3^{\text{PI}}=\frac{G_F^2m_c^2}{6\pi}\sum\limits_{d'=d,s}\sum\limits_{q=d,s}|V_{cq}|^2|V_{ud'}|^2\left(F^{qu}Q^{d'}+F_S^{qu}Q_S^{d'}+G^{qu}T^{d'}+G_S^{qu}T_S^{d'}\right),\\
&\mathcal{T}_3^{\text{WA}_0}=\frac{G_F^2m_c^2}{6\pi}\sum\limits_{q=d,s}\sum\limits_{q'=d,s}|V_{cq}|^2|V_{uq'}|^2\left(F^{qq'}Q^u+F_S^{qq'}Q_S^u+G^{qq'}T^u+G_S^{qq'}T_S^u\right),\\
&\mathcal{T}_3^{\text{WA}_+}=\frac{G_F^2m_c^2}{6\pi}\sum\limits_{s'=d,s}|V_{cs'}|^2\Bigg[\sum\limits_{q'=d,s}|V_{uq'}|^2\left(\tilde{F}^{uq'}Q^{s'}+\tilde{F}_S^{uq'}Q_S^{s'}+\tilde{G}^{uq'}T^{s'}+\tilde{G}_S^{uq'}T_S^{s'}\right)\\
&\hspace{4.3cm}+\sum\limits_{l=e,\mu}\left(\tilde{F}^{\nu l}Q^{s'}+\tilde{F}_S^{\nu l}Q_S^{s'}+\tilde{G}^{\nu l}T^{s'}+\tilde{G}_S^{\nu l}T_S^{s'}\right)\Bigg].
\end{aligned}
\label{eq:Ts}
\end{equation}
The label \(qq'\) in \(F^{qq'},\dots,G_S^{qq'}\) refers to the flavors of the quarks in the \(qq'\) loop in Fig. \ref{fig:Gamma3}.
The Wilson coefficients \(F,G\) are functions of the mass ratio \(z=m_s^2/m_c^2\) and \(\mu_0/m_c\), where \(\mu_0\) denotes the renormalization scale for
\(\Delta C=0\) operators. These dimension-6 operators read as follows:
\begin{equation}
\begin{aligned}
Q^q=\bar{c}\gamma_\mu(1-\gamma_5)q\ \bar{q}\gamma^\mu(1-\gamma_5)c,\hspace{1.9cm} &Q_S^q=\bar{c}(1-\gamma_5)q\ \bar{q}(1+\gamma_5)c,\\
T^q=\bar{c}\gamma_\mu(1-\gamma_5)T^aq\ \bar{q}\gamma^\mu(1-\gamma_5)T^ac,\hspace{1cm} &T_S^q=\bar{c}(1-\gamma_5)T^aq\ \bar{q}(1+\gamma_5)T^ac.
\end{aligned}
\label{eq:DeltaC0ops}
\end{equation}
The LO Wilson coefficients \(F^{qq'},\dots,G_S^{qq'}\) for $B$ mesons can be found in Refs. \cite{uraltsev96,neubert96}. The NLO QCD corrections have been computed in Refs. \cite{beneke02,ROME}. 
The Wilson coefficients \(\tilde{F}^{qq'},\dots,\tilde{G}_S^{qq'}\) for \(\text{WA}_+\) have been calculated at LO QCD for the \(B_c\) meson in Ref. \cite{beneke96b}.
The NLO corrections for the nonleptonic \(\text{WA}_+\) can be determined from the published results via a Fierz transformation of the \(\Delta C=1\)
operators given in Eq. \eqref{eq:DeltaC1ops}. With our choice of evanescent operators \cite{buras90}, the Fierz symmetry is respected at the one-loop level. This allows us to obtain
the following relation between Wilson coefficients, that holds up to NLO:
\begin{equation}
 \left(\tilde{F}^{ud},\tilde{F}_S^{ud},\tilde{G}^{ud},\tilde{G}_S^{ud}\right)=\left(F^{sd},F_S^{sd},G^{sd},G_S^{sd}\right)(C_1\lra C_2,m_s=0),
\end{equation}
where \(C_1,C_2\) are the Wilson coefficients of the respective \(\Delta C=1\) operator.
The NLO coefficients for the semileptonic weak annihilation \(\tilde{F}^{\nu l},\dots,\tilde{G}_S^{\nu l}\) have been computed for the first time
and are given in Appendix \ref{sec:SLWA}.\\
The subleading \(1/m_c^4\) contribution of the HQE is expected to be sizeable in the charm sector. It furthermore provides a crucial test of the convergence properties of the expansion.
This contribution is the leading correction in an expansion of the spectator effects in the momentum and mass of the spectator quark.
Applying the same decomposition as for \(\mathcal{T}_3\), we find
\begin{equation}
\begin{aligned}
&\mathcal{T}_4^{\text{PI}}=\frac{G_F^2m_c^2}{6\pi}\sum\limits_{d'=d,s}\sum\limits_{q=d,s}|V_{cq}|^2|V_{ud'}|^2\sum\limits_{i=1}^6\left(g_i^{qu}P_i^{d'}+h_i^{qu}S_i^{d'}\right),\\
&\mathcal{T}_4^{\text{WA}_0}=\frac{G_F^2m_c^2}{6\pi}\sum\limits_{q=d,s}\sum\limits_{q'=d,s}|V_{cq}|^2|V_{uq'}|^2\sum\limits_{i=1}^6\left(g_i^{qq'}P_i^u+h_i^{qq'}S_i^u\right),\\
&\mathcal{T}_4^{\text{WA}_+}=\frac{G_F^2m_c^2}{6\pi}\sum\limits_{s'=d,s}|V_{cs'}|^2\sum\limits_{i=1}^6\Bigg[\sum\limits_{q=d,s}|V_{uq'}|^2\left(\tilde{g}_i^{uq'}P_i^{s'}+\tilde{h}_i^{uq'}S_i^{s'}\right)\\
&\hspace{4.3cm}+\sum\limits_{l=e,\mu}\left(\tilde{g}_i^{\nu l}P_i^{s'}+\tilde{h}_i^{\nu l}S_i^{s'}\right)\Bigg].
\end{aligned}
\label{eq:T4s}
\end{equation}
The dimension-7 operators and Wilson coefficients are given in Appendix \ref{sec:dim7}. For the case of QCD operators (see the discussion at the end
of this section) this contribution has previously been determined in Ref. \cite{gabbiani03} using a different operator basis.\\
\begin{figure}[t]
\centering
  \includegraphics[width=0.3\textwidth]{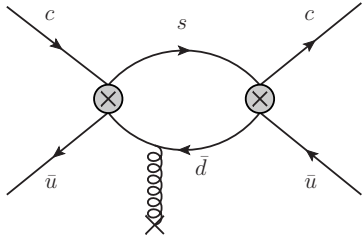}
\caption{Sample diagram for the chromomagnetic contribution to the spectator corrections of order \(1/m_c^5\) in the HQE.\label{fig:dimension8}}
\end{figure}
A comment about the \(\mathcal{T}_5\) term is in order. In addition to the kinetic corrections, there is also a chromomagnetic contribution to the spectator effects. The kinetic corrections
can be computed in the same fashion as for \(\mathcal{T}_4\) and are found to be numerically unimportant. The chromomagnetic effects of the form shown in Fig. \ref{fig:dimension8}
can, however, not be estimated in the vacuum saturation approximation (VSA) \cite{VSA}.
But if these contributions are not severely enhanced compared to the kinetic effects, the HQE can be truncated to good approximation after the \(\mathcal{T}_4\) term.\\
As stressed in Refs. \cite{ROME,becirevic01}, the two possible contractions shown in in Fig. \ref{fig:contractions} have to be considered when computing the matrix elements of the dimension-6 operators
\(\left(\bar{c}\Gamma_iq\right)\left(\bar{q}\Gamma_i'c\right)\) on the lattice. The eye contraction diagram induces mixing of the renormalized dimension-6 operators into lower-dimensional
operators. The required power subtraction of this mixing poses considerable challenges for lattice computations. We therefore distinguish between these two contributions in our parametrization
of hadronic matrix elements in Appendix \ref{sec:parametrization}.
Such mixing also occurs in $\mathcal{T}^{\text{sing}}$ at $\mathcal{O}(\alpha_s)$ in perturbation theory. For dimensional reasons, this mixing has to
be of the form \(\propto \alpha_s(m_c)m_c^3(\bar{c}c)\). In QCD, a perturbative subtraction of this term is necessary as discussed in Ref. \cite{beneke98}.
A NLO computation of $\mathcal{T}^{\text{sing}}$ has not been performed so far and is also beyond the scope of this work. Yet, in the ratio $\tau(D^+)/\tau(D^0)$,
the contribution of $\mathcal{T}^{\text{sing}}$ and the eye contraction cancel due to isospin symmetry. Unfortunately, the deviations from exact \(SU(3)\)
flavor symmetry are too large in \(\tau(D_s^+)/\tau(D^0)\) to be ignored. For the analysis of \(\tau(D_s^+)/\tau(D^0)\), we thus match the QCD operators
to HQET operators, where the natural cutoff due to the limit \(m_c\ra\infty\) guarantees the absence of mixing with lower-dimensional
operators in perturbation theory \cite{beneke98,ROME,becirevic01}.
A HQET description of operators, however, affects the subleading \(1/m_c\) corrections, because the QCD operators \(\bar{c}\Gamma q\ \bar{q}\Gamma c\)
coincide with the respective HQET operators only up to \(1/m_c\) corrections. The expansion of the QCD operators in HQET yields
\begin{equation}
 \bar{c}\Gamma q\ \bar{q}\Gamma c=\bar{h}_v\Gamma q\ \bar{q}\Gamma h_v+\frac{1}{2m_c}\left[\bar{h}_v\left(-i\overleftarrow{\slashed{D}}\right)\Gamma q\ \bar{q}\Gamma h_v
+\bar{h}_v\Gamma q\ \bar{q}\Gamma\left(i\slashed{D}\right)h_v\right]+\mathcal{O}\left(\frac{1}{m_c^2}\right).
\end{equation}
The operators arising this way are \(P_{5,6}^q\) and \(S_{5,6}^q\) in Eq. \eqref{eq:dim7ops}. The respective terms are absent in QCD when the hadronic matrix elements are
determined to all orders in \(1/m_c\).
%
%
%
%
%
\section{Phenomenology\label{sec:phenomenology}}
We perform an analysis of the lifetime ratios \(\tau(D^+)/\tau(D^0)\) and \(\tau(D_s^+)/\tau(D^0)\).
Since the pole mass definition contains an infrared renormalon ambiguity \cite{beneke94,bigi94m}, we use the \(\overline{\text{MS}}\) in
addition to the pole mass scheme. In the \(\overline{\text{MS}}\) scheme, we use \(\overline{z}=\overline{m}_s^2(m_c)/\overline{m}_c^2(m_c)\).
As discussed in detail in Ref. \cite{beneke02}, this sums up terms of the form \(\alpha_s^n(\mu_1)z\ln^nz\) to all orders in perturbation theory.
\subsection{The ratio \(\tau(D^+)/\tau(D^0)\)\label{sec:DpD0}}
\begin{figure}[t]
\centering
\subfloat[Pole mass scheme]{
        \includegraphics[width=0.4\textwidth]{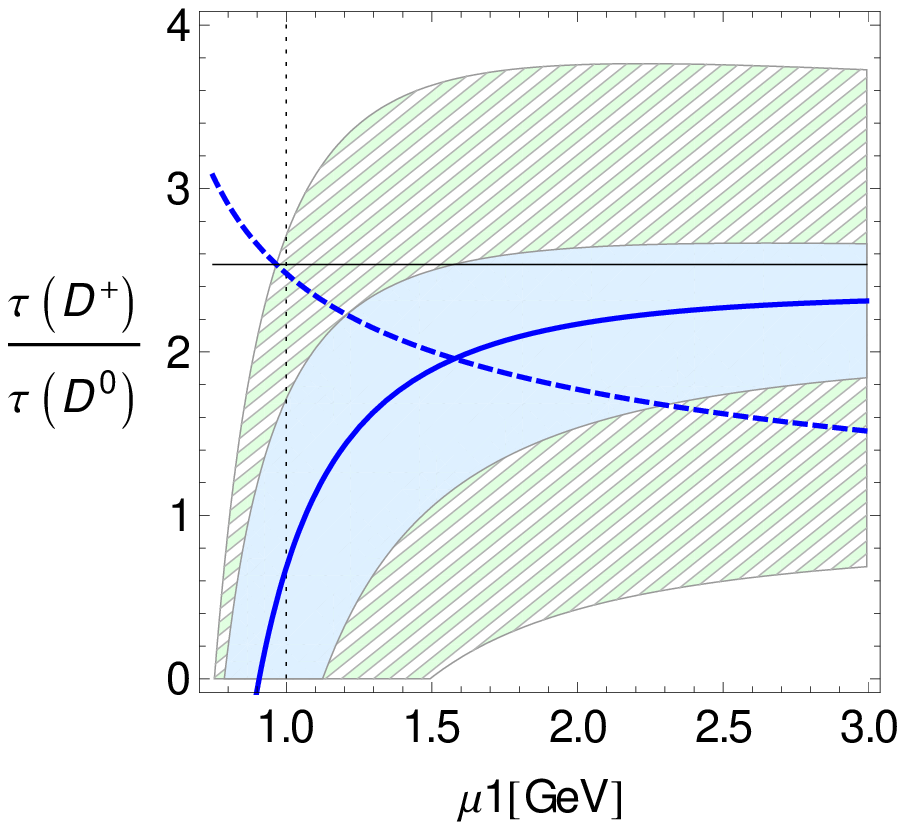}
        \label{fig:DpD0pole}
}
\quad
\subfloat[\(\overline{\text{MS}}\) mass scheme]{
        \includegraphics[width=0.4\textwidth]{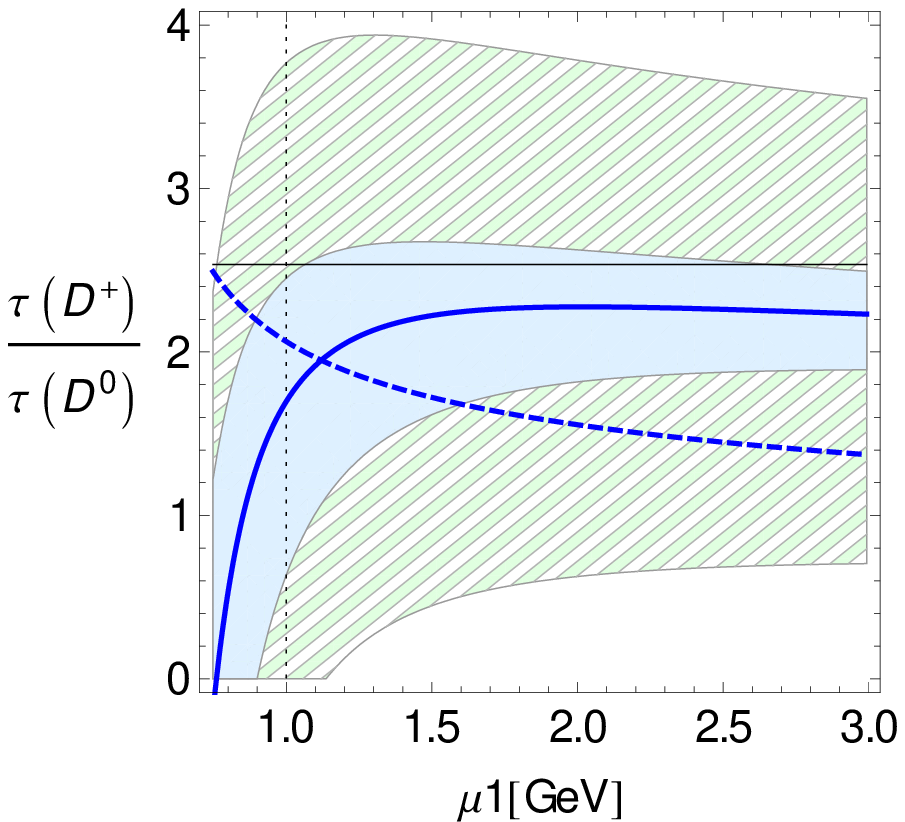}
        \label{fig:DpD0MS}
    }
\caption{\(\tau(D^+)/\tau(D^0)\) in the pole and \(\overline{\text{MS}}\) mass scheme plotted over the scale \(\mu_1\). The black horizontal
line shows the experimental value, and the solid (dashed) line the NLO (LO) prediction using the matrix elements extracted from Ref. \cite{becirevic01}.
The hatched and shaded regions show the theoretical uncertainties for the VSA and the extracted values, respectively. The dotted vertical line
marks the lower limit \(\mu_1=1\text{ GeV}\) of the region in which we vary the renormalization scale for the numerical evaluations.\label{fig:DpD0QCD}}
\end{figure}

We determine the ratio \(\tau(D^+)/\tau(D^0)\), first using QCD operators, and then briefly discuss the HQET case. Isospin symmetry implies the following relations:
\begin{equation}
\begin{aligned}
 \frac{\Braket{D^0|\left(\vec{Q},\vec{P},\vec{S}\right)^{u,d}|D^0}}{2M_{D^0}}&=\frac{\Braket{D^+|\left(\vec{Q},\vec{P},\vec{S}\right)^{d,u}|D^+}}{2M_{D^+}},\\
 \frac{\Braket{D^0|\left(\vec{Q},\vec{P},\vec{S}\right)^{s}|D^0}}{2M_{D^0}}&=\frac{\Braket{D^+|\left(\vec{Q},\vec{P},\vec{S}\right)^{s}|D^+}}{2M_{D^+}}.
\end{aligned}
 \end{equation}
From Eqs. \eqref{eq:Ts}, \eqref{eq:T4s} and \eqref{eq:QCDparametrization},\eqref{eq:QCDparametrization7}, we obtain \(\tau(D^+)/\tau(D^0)\):
\begin{equation}
 \begin{aligned}
 \Gamma(D^0)-\Gamma(D^+)=\frac{G_F^2m_c^2}{12\pi}f_D^2M_D\Bigg[&\left(\vec{F}^{s'd'}-|V_{ud}|^2\vec{F}^{s'u}\right)\cdot\vec{B}\\
 &+\frac{M_D-m_c}{m_c}\left(\vec{f}^{s'd'}-|V_{ud}|^2\vec{f}^{s'u}\right)\cdot\vec{b}\Bigg].
 \end{aligned}
\label{eq:Gamma0mGammap}
\end{equation}
For brevity, we have introduced the vector notation
\begin{equation}
 \vec{F}^{qq'}=\left(\begin{array}{c}F^{qq'}\\F_S^{qq'}\\G^{qq'}\\G_S^{qq'}\end{array}\right), \hspace{0.3cm} \vec{B}=\left(\begin{array}{c}B_1\\B_2\\\epsilon_1\\\epsilon_2\end{array}\right), \hspace{0.3cm}
 \vec{f}^{qq'}=\left(\begin{array}{c}g_3^{qq'}\\g_4^{qq'}\\h_3^{qq'}\\h_4^{qq'}\end{array}\right), \hspace{0.3cm} \vec{b}=\left(\begin{array}{c}-\rho_3\\ \rho_4\\ -\sigma_3\\ \sigma4\end{array}\right).
\end{equation}
The NLO QCD correction to \(\vec{F}^{ss}\) has not been determined. Following Ref. \cite{beneke02}, we thus set \(|V_{ud}|^2=1\) and \(V_{us}=0\) in the NLO term. The induced error is of order 
\(|V_{us}|^2\alpha_s(m_c)z\log{z}\), which is of order \(10^{-3}\) and thus negligible. Furthermore, the Cabibbo and chirality-suppressed weak annihilation contribution to \(D^+\) is neglected.
The matrix elements of the \(\Delta C=1\) operators can be estimated within the VSA \cite{VSA}. The uncertainties are expected to be of order \(1/N_C\), although calculations in the $B$ sector
\cite{baek97,pierro98,becirevic01} hint at much smaller errors for the color octet operators. Thus, using
\begin{equation}
\begin{aligned}
 &\left(B_1,B_2,\epsilon_1,\epsilon_2\right)=\left(1\pm\frac13,\left(1+2\frac{M_D-m_c}{m_c}\right)\left(1\pm\frac13\right),0\pm\frac{1}{10},0\pm\frac{1}{10}\left(1+2\frac{M_D-m_c}{m_c}\right)\right),\\
 &\left(\rho_3,\rho_4,\sigma_3,\sigma_4\right)=\left(1\pm\frac13,1\pm\frac13,0\pm\frac{1}{10},0\pm\frac{1}{10}\right),\\
\end{aligned}
\end{equation}
we obtain in the pole and \(\overline{\text{MS}}\) mass schemes with the input parameters given in Tab. \ref{tab:expinputs}:
\begin{equation}
 \begin{aligned}
  &\left(\frac{\tau(D^+)}{\tau(D^0)}\right)_\text{exp}\hspace{0.7cm}=2.536\pm0.019,\\
  &\left(\frac{\tau(D^+)}{\tau(D^0)}\right)_\text{pole,VSA}=1.9\pm1.7^{(\text{hadronic})\hspace{0.15cm}+0.6\hspace{0.1cm}(\text{scale})}_{\hspace{1.5cm}-1.5}\pm0.0^{(\text{parametric})},\\
  &\left(\frac{\tau(D^+)}{\tau(D^0)}\right)_{\overline{\text{MS}}\text{,VSA}}\hspace{0.08cm}=2.2\pm1.7^{(\text{hadronic})\hspace{0.15cm}+0.3\hspace{0.1cm}(\text{scale})}_{\hspace{1.5cm}-0.7}\pm0.1^{(\text{parametric})}.
 \end{aligned}
 \label{eq:DpD0QCDVSAresult}
\end{equation}
We have varied \(\mu_0\) and \(\mu_1\) from \(1\text{ GeV}\) to \(2m_c\). We do not use the full region \(0.5 m_Q -2m_Q\) common in $B$ decays, because we
do not trust perturbation theory to hold below about \(1\text{ GeV}\). The overall error is largely driven by hadronic uncertainties.
The size of the subleading \(1/m_c\) corrections relative to the leading spectator effects is an important check on the convergence behavior. We find that
\begin{equation}
 \frac{\Gamma_4^{(0)}(D^0)-\Gamma_4^{(0)}(D^+)}{\Gamma_3^{(0)}(D^0)-\Gamma_3^{(0)}(D^+)}\approx-50\%,
\end{equation}
which is large, but compatible with a convergent series. The \(1/m_c^5\) term should be numerically less relevant, as discussed in Sec. \ref{sec:HQE}. Next-to-leading-order QCD corrections
to \(\mathcal{T}_3\) are at a level of below \(30\%\) near the charm scale.\\
The predictive power of the VSA is very limited. In the following, we perform a very aggressive estimation of \(\tau(D^+)/\tau(D^0)\)
by extracting the bag parameters from a lattice calculation in the $B$ sector \cite{becirevic01} and ignore any possible systematic uncertainties related with this
approach. We extract the bag parameters for a meson mass of \(m_P=1.8\text{ GeV}\) and a hadronic scale \(\mu_0=2.7 \text{ GeV}\) from Ref. \cite{becirevic01} and evaluate
this to the charm scale \(\mu_0=m_c\) at NLO. The required anomalous dimension matrices can be inferred from Ref. \cite{ADM}. This reduces the hadronic uncertainty considerably:
\begin{equation}
 \begin{aligned}
  &\left(\frac{\tau(D^+)}{\tau(D^0)}\right)_\text{pole, extr. from \cite{becirevic01}}=1.9\pm0.5^{(\text{hadronic})\hspace{0.15cm}+0.6\hspace{0.1cm}(\text{scale})}_{\hspace{1.5cm}-1.4}\pm0.0^{(\text{parametric})},\\
  &\left(\frac{\tau(D^+)}{\tau(D^0)}\right)_{\overline{\text{MS}}\text{, extr. from \cite{becirevic01}}}\hspace{0.08cm}=2.2\pm0.4^{(\text{hadronic})\hspace{0.15cm}+0.3\hspace{0.1cm}(\text{scale})}_{\hspace{1.5cm}-0.7}\pm0.0^{(\text{parametric})}.
 \end{aligned}
 \label{eq:DpD0QCDBecresult}
\end{equation}
The dependence on the renormalization scale of the \(\Delta C=1\) operators is illustrated in Fig. \ref{fig:DpD0QCD}. It is dominated by the scale dependence
of the subleading dimension-7 contributions, because they are only evaluated at LO QCD. Also, the difference between the expectation values in the
pole mass and \(\overline{\text{MS}}\) mass scheme originates dominantly from \(\Gamma_4\). The \(\mu_1\) dependence of Fig. \ref{fig:DpD0QCD} suggests that
perturbation theory becomes unreliable at about \(1\text{ GeV}\), but it seems to be under control at the charm threshold. We see a substantial reduction of the
theoretical uncertainties from the VSA to the extracted matrix elements. The equality of the central values is coincidental.

In HQET, we get an expression similar to Eq. \eqref{eq:Gamma0mGammap} for \(\tau(D^+)/\tau(D^0)\):
\begin{equation}
 \begin{aligned}
 \Gamma(D^0)&-\Gamma(D^+)=\frac{G_F^2m_c^2}{12\pi}f_D^2M_D\Bigg[\left(\vec{F}^{s'd'}-|V_{ud}|^2\vec{F}^{s'u}-|V_{cd}|^2\left(\vec{F}^{ud'}+\vec{F}^{\nu e}+\vec{F}^{\nu\mu}\right)\right)\cdot\vec{B}\\
 &+\frac{M_D-m_c}{m_c}\sum\limits_{j=3}^6(-1)^j\Big[\left(g_j^{s'd'}-|V_{ud}|^2g_j^{s'u}-|V_{cd}|^2\left(g_j^{ud'}+g_j^{\nu e}+g_j^{\nu\mu}\right)\right)\rho_j\\
 &\hspace{4cm}+\left(g_j\ra h_j,\rho_j\ra\sigma_j\right)\Big]\Bigg],
 \end{aligned}
\label{eq:Gamma0mGammapHQET}
\end{equation}
where \(\vec{F}\) and \(\vec{B}\) are defined as before and we have set \(m_u=m_d=0\).
The non-Cabibbo-suppressed \(\delta\)'s cancel in the difference \(|V_{ud}|^2\left(\vec{F}^{sd}-\vec{F}^{su}\right)\) because of isospin symmetry.
We neglect the remaining ones because of Cabibbo suppression. In the VSA, we obtain
\begin{equation}
 \begin{aligned}
  &\left(\frac{\tau(D^+)}{\tau(D^0)}\right)_\text{pole,VSA}=2.4\pm2.0^{(\text{hadronic})\hspace{0.15cm}+0.3\hspace{0.1cm}(\text{scale})}_{\hspace{1.5cm}-0.8}\pm0.0^{(\text{parametric})},\\
  &\left(\frac{\tau(D^+)}{\tau(D^0)}\right)_{\overline{\text{MS}}\text{,VSA}}\hspace{0.08cm}=2.6\pm1.9^{(\text{hadronic})\hspace{0.15cm}+0.2\hspace{0.1cm}(\text{scale})}_{\hspace{1.5cm}-0.5}\pm0.1^{(\text{parametric})}.
 \end{aligned}
 \label{eq:DpD0HQETVSAresult}
\end{equation}
The sizeable differences between the HQET and the QCD results in the VSA seem puzzling at first, but we have to remember that the matrix elements are
defined in a different scheme. The transformation law for the dimension-6 Wilson coefficients is given in Refs. \cite{beneke02,ROME}.
We have checked explicitly that this relation holds for our numerical coefficients, if in HQET we neglect weak annihilation in \(D^+\)
and set \(|V_{ud}|^2=1,\hspace{0.05cm}V_{us}=0\) at NLO as we have in QCD. This scheme dependence is canceled by the scheme dependence of the operators.
The VSA is, however, not sensitive to the scheme, and the numerical deviation between Eqs. \eqref{eq:DpD0QCDVSAresult} and \eqref{eq:DpD0HQETVSAresult} is just a consequence
of this. This once more emphasizes the dire need for lattice inputs for the matrix elements.
%
%
%
%
\subsection{The ratio \(\tau(D_s^+)/\tau(D^0)\)\label{sec:DspD0}}
Since \(SU(3)\) flavor symmetry is rather crude in the case of \(\tau(D_s^+)/\tau(D^0)\), the contributions of $\mathcal{T}^{\text{sing}}$ and the eye contraction do not fully cancel
as was the case in \(\tau(D^+)/\tau(D^0)\). We hence use only HQET operators in the following analysis.
The dominant sources for the lifetime difference between these mesons have been identified in Ref. \cite{bigi94}.
We further include (e), which could possibly contribute at the level of a few percent.
\begin{compactenum}[(a)]
 \item The decay \(D_s^+\ra\tau^+\nu\).
 \item \(SU(3)\) flavor breaking in \(\mu_\pi^2\) and \(\mu_G^2\).
 \item The weak annihilation in \(D^0\) and \(D_s^+\).
 \item The Cabibbo-suppressed Pauli interference in \(D_s^+\).
 \item \(SU(3)\) flavor breaking in the nonvalence part of the \(c\bar{d}\) Pauli interference.
\end{compactenum}
The first effect (a) cannot be properly dealt with in the HQE, because the energy release in \(D_s^+\ra\tau^+\nu\) is just \(\sim200\text{ MeV}\).
Instead, we define a subtracted \(D_s^+\) lifetime by
\begin{equation}
 \overline{\tau}(D_s^+)=\frac{\tau(D_s^+)}{1-\text{Br}(D_s^+\ra\tau^+\nu)}=\left(529\pm8\right)\cdot10^{-15}\mbox{ s}
\end{equation}
and compare our prediction with \(\overline{\tau}(D_s^+)/\tau(D^0)\).\\
\(SU(3)\) flavor breaking in \(\tau\left(D_s^+\right)/\tau\left(D^0\right)\) arises at order \(\left(\Lambda\middle/m_c\right)^2\) in the HQE.
We follow Ref. \cite{bigi11} to extract the corresponding matrix elements of the dimension-5 operators from experimental data.
The expectation value \(\mu_G^2\) of the chromomagnetic operator can be extracted from the hyperfine splitting. We find, using the meson masses given in Ref. \cite{pdg},
\begin{equation}
 \frac{\mu_G^2(D_s^+)}{\mu_G^2(D^0)}\simeq\frac{M_{D_s^{+*}}-M_{D_s^+}}{M_{D^{0*}}-M_D^0}=1.012\pm0.003.
\label{eq:muGDsD0}
\end{equation}
The effects of \(1/m_c\) corrections should cancel to a large extent in the ratio in Eq. \eqref{eq:muGDsD0}. Regarding the overall uncertainties, this effect can safely be neglected.
The situation in the case of the kinetic operator is less clear. Yet we can estimate the difference \(\mu_\pi^2(D_s)-\mu_\pi^2(D)\) from spectroscopy. We obtain
\begin{equation}
\begin{aligned}
 \mu_\pi^2(D_s)-\mu_\pi^2(D^0)\simeq &\frac{2m_bm_c}{m_b-m_c}\left[\left(\langle M_{D_s^+}\rangle-\langle M_{D^0}\rangle\right)-
\left(\langle M_{B_s^0}\rangle-\langle M_{B^+}\rangle\right)\right]\\
&=(0.07\pm0.01)\mbox{ GeV}^2,
\end{aligned}
\label{eq:mupiextraction}
\end{equation}
where
\begin{equation}
 \langle M_{D}\rangle=\frac14\left(M_D+3M_{D^*}\right).
\end{equation}
Equation \eqref{eq:mupiextraction} holds up to relative \(1/m_c\) and \(1/m_b\) corrections, which do not cancel here. Including the higher-order effects, we expect up to \cite{bigi11}
\begin{equation}
 \mu_\pi^2(D_s^+)-\mu_\pi^2(D^0)\sim0.1\mbox{ GeV}^2,
\end{equation}
which corresponds to about \(25\%\) \(SU(3)\) flavor breaking in \(\mu_\pi^2\). Fortunately,  this effect can be included at NLO independent of the coefficients \(c_5^f\):
\begin{equation}
\begin{aligned}
 \left(\frac{\overline{\tau}(D_s^+)}{\tau(D^0)}-1\right)_{(\text{b})}=&\frac{G_F^2m_c^5}{192\pi^3}\sum\limits_f|V_{\text{CKM}}|^2\Bigg[c_3^f\frac{\mu_\pi^2(D_s^+)-\mu_\pi^2(D^0)}{2m_c^2}\\
&+(c_3^f+4c_5^f)\frac{\mu_G^2(D^0)-\mu_G^2(D_s^+)}{2m_c^2}+\mathcal{O}\left(\frac{1}{m_c^3}\right)\Bigg]\cdot\overline{\tau}(D_s^+)
\end{aligned}
\end{equation}
Numerically, we find
\begin{equation}
 \left(\frac{\overline{\tau}(D_s^+)}{\tau(D^0)}-1\right)_\text{(b)}=
 \begin{cases}
   \left(0.19_{-0.03}^{+0.04}\right)\frac{\mu_\pi^2(D_s^+)-\mu_\pi^2(D^0)}{\text{GeV}^2},&\text{pole mass scheme}\\
   \left(0.16_{-0.02}^{+0.03}\right)\frac{\mu_\pi^2(D_s^+)-\mu_\pi^2(D^0)}{\text{GeV}^2},&\overline{\text{MS}}\text{ mass scheme}\\
  \end{cases}
  \label{eq:DspD0b}
\end{equation}
which enhances \(\overline{\tau}(D_s^+)/\tau(D^0)\) by \(2\%\) for \(\mu_\pi^2(D_s^+)-\mu_\pi^2(D^0)=0.1\text{ GeV}^2\).\\
The weak annihilation effects (c) are Cabibbo leading, but do suffer from chirality suppression. Chirality breaking stems from final-state masses and
QCD effects. Since the mass ratio \(z=m_s^2/m_c^2\) is rather small, the NLO corrections to the Wilson coefficients are very important here to obtain
a meaningful result. The weak annihilation contributions are given by
\begin{equation}
 \begin{aligned}
 &\left[\Gamma(D^0)-\Gamma(D_s^+)\right]_{\text{WA}(D_s^+)}=-\frac{G_F^2m_c^2}{12\pi}f_{D_s}^2M_{D_s}|V_{cs}|^2\Bigg[\left(\vec{\tilde{F}}^{ud'}+\vec{\tilde{F}}^{\nu e}+\vec{\tilde{F}}^{\nu\mu}\right)
 \cdot\left(\vec{B}^s+\Delta\vec{\delta}\right)\\
 &+\sum\limits_{j=1}^6M_j\left[\left(g_j^{s'u}+g_j^{\nu e}+g_j^{\nu\mu}\right)\left(\rho_j^s+\Delta\delta_{\rho,j}\right)
 +\left(h_j^{s'u}+h_j^{\nu e}+h_j^{\nu\mu}\right)\left(\sigma_j^s+\Delta\delta_{\sigma,j}\right)\right]\Bigg]
 \end{aligned}
\label{eq:DspWA}
\end{equation}
for \(D_s^+\) and
\begin{equation}
 \begin{aligned}
 \left[\Gamma(D^0)-\Gamma(D_s^+)\right]_{\text{WA}(D^0)}=&\frac{G_F^2m_c^2}{12\pi}f_{D}^2M_{D}\Bigg[\vec{\tilde{F}}^{s'd'}\cdot\left(\vec{B}^u+\tilde{\Delta}\vec{\delta}\right)\\
 &+\sum\limits_{j=1}^6M_j\left[g_j^{s'd'}\left(\rho_j^u+\tilde{\Delta}\delta_{\rho,j}\right)+h_j^{s'd'}\left(\sigma_j^u+\tilde{\Delta}\delta_{\sigma,j}\right)\right]\Bigg]
 \end{aligned}
\label{eq:D0WA}
\end{equation}
for \(D^0\). We have introduced the following notation:
\begin{equation}
 M_j=\begin{cases}
      -\frac{m_s}{m_q},&j=1,2,\\
      (-1)^j\frac{M_{D_s}-m_c}{m_c},&j=3,4,\\
      (-1)^j\frac{M_{D_s}-m_c-m_s}{m_c},&j=5,6.
     \end{cases}
\end{equation}
The \(\delta_i^{qq'}\) only enter in the \(SU(3)\)-breaking combinations
\begin{equation}
\begin{aligned}
 \Delta\vec{\delta}_{(\rho,\sigma)}=\vec{\delta}_{(\rho,\sigma)}^{ss}-\frac{f_D^2M_D}{f_{D_s}^2M_{D_s}}\vec{\delta}_{(\rho,\sigma)}^{us},\hspace{1cm}
 \tilde{\Delta}\vec{\delta}_{(\rho,\sigma)}=\vec{\delta}_{(\rho,\sigma)}^{uu}-\frac{f_{D_s}^2M_{D_s}}{f_D^2M_D}\vec{\delta}_{(\rho,\sigma)}^{su}.
\end{aligned}
\end{equation}
These weak annihilation contributions depend strongly on the amount of chirality breaking through the matrix elements. Here, the VSA is far too crude, and we thus estimate
this using experimental results for semileptonic rates similar to the study in Ref. \cite{gambino10}. They allow a very clean extraction of the chirality-breaking combinations
\(B_1-B_2\) and \(\epsilon_1-\epsilon_2\). The experimental average for the ratios of semileptonic rates is
\begin{equation}
\left. \frac{\Gamma(D_s^+\ra Xe^+\nu)}{\Gamma(D^0\ra Xe^+\nu)}\right|_{\text{\cite{pdg}}}=0.821\pm0.054.
\end{equation}
The difference of the semileptonic rates arises first at order \(1/m_c^2\) in the HQE because of \(SU(3)\) flavor breaking.
The dominant effect, however, is due to the semileptonic weak annihilation at order \(1/m_c^3\) and higher in the HQE. In terms of
the required matrix elements, we obtain in the \(\overline{\text{MS}}\) mass scheme
\begin{equation}
 \begin{aligned}
  \left.\frac{\Gamma(D_s^+\ra Xe^+\nu)}{\Gamma(D^0\ra Xe^+\nu)}\right|_{\overline{\text{MS}}}=&1+\left(0.25\pm0.03\right)\frac{\mu_\pi^2(D_s^+)-\mu_\pi^2(D^0)}{\text{GeV}^2}\\
  &-\left(7.10\pm1.23\right)(B_1^s+\Delta\delta_1-B_2^s-\Delta\delta_2)-\left(2.56\pm1.10\right)\left(\epsilon_1^s+\Delta\delta_3\right)\\
  &+\left(2.38\pm1.05\right)\left(\epsilon_2^s+\Delta\delta_4\right)+0.51\rho_1^s+0.51\rho_2^s-4.48\rho_3^s+1.73\rho_5^s+1.73\rho_6^s.
 \end{aligned}
\end{equation}
\begin{figure}[t]
\centering
\subfloat[Pole mass scheme]{
        \includegraphics[width=0.35\textwidth]{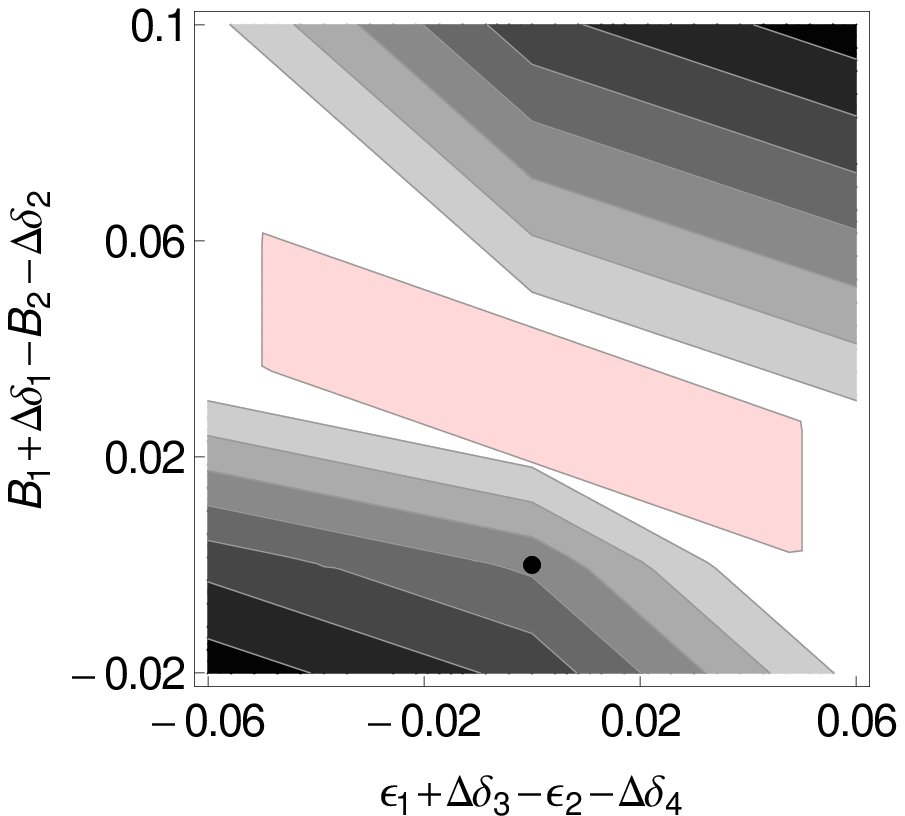}
        \label{fig:SLExtractionpole}
}
\quad
\subfloat[\(\overline{\text{MS}}\) mass scheme]{
        \includegraphics[width=0.35\textwidth]{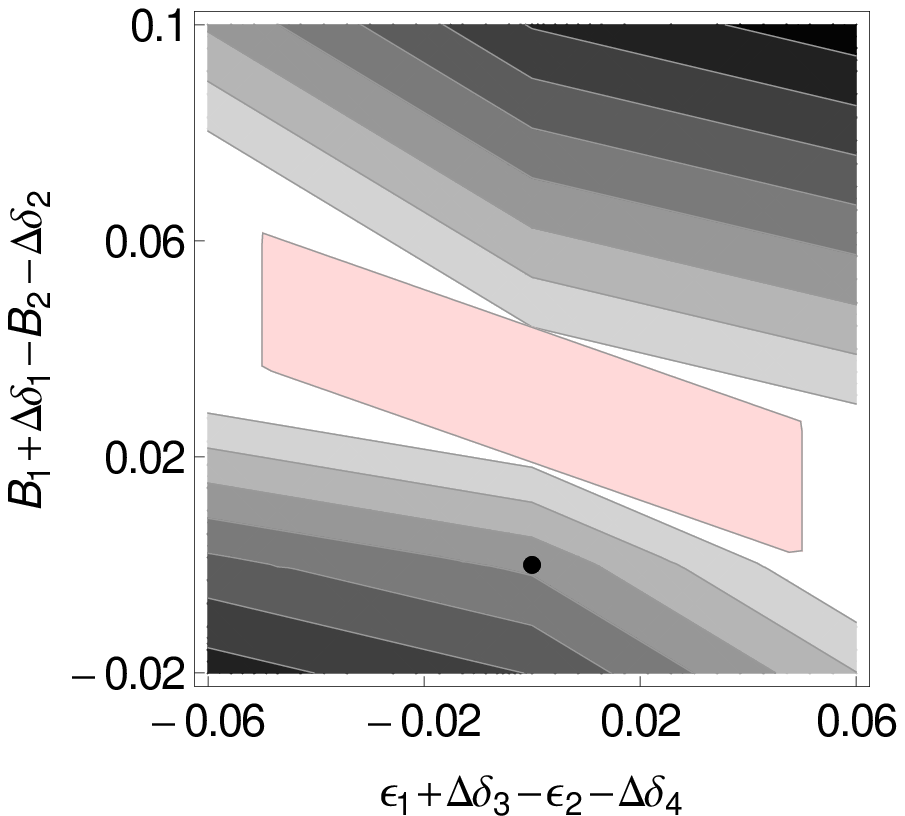}
        \label{fig:SLExtractionMS}
    }
\caption{Contour plot of \(\left.\left|\left(\frac{\Gamma_{\text{sl}}(D_s^+)}{\Gamma_{\text{sl}}(D^0)}\right)_{\text{theory}}
-\left(\frac{\Gamma_{\text{sl}}(D_s^+)}{\Gamma_{\text{sl}}(D^0)}\right)_{\text{exp}}\right|\middle/\left(\Delta\frac{\Gamma_{\text{sl}}(D_s^+)}{\Gamma_{\text{sl}}(D^0)}\right)_{\text{exp}}\right.\)
over the chirality-breaking combinations \(\epsilon_1^s+\Delta\delta_3-\epsilon_2^s-\Delta\delta_4\) and \(B_1^s+\Delta\delta_1-B_2^s-\Delta\delta_2\) in the pole and \(\overline{\text{MS}}\) mass schemes.
The contours correspond to the \(1,2,\dots\hspace{0.05cm}\sigma\) regions. The black dot marks the VSA point. We used \(\Delta\mu_\pi^2\equiv\mu_\pi^2(D_s^+)-\mu_\pi^2(D^0)=0.1\) in the
\(\mathcal{T}_2\) contribution. The red region indicates the matrix element space we use for further evaluation of \(\tau(D_s^+)/\tau(D^0)\).\label{fig:SLExtraction}}
\end{figure}
Setting all \(\rho_i^s\) equal to 1 (VSA), we obtain the constraints illustrated in Fig. \ref{fig:SLExtraction}
\begin{equation}
\begin{aligned}
  &B_1^s+\Delta\delta_1-B_2^s-\Delta\delta_2=0.032-0.350\left(\epsilon_1^s+\Delta\delta_3-\epsilon_2^s-\Delta\delta_4\right)\pm0.013,\\
  &\frac{B_1+\Delta\delta_1+B_2+\Delta\delta_2}{2}=1\pm1/3,\\
  &\epsilon_1^s+\Delta\delta_3-\epsilon_2^s-\Delta\delta_4=0\pm0.05,\\
  &\epsilon_1^s+\Delta\delta_3+\epsilon_2^s+\Delta\delta_4=0\pm0.05.
\end{aligned}
\end{equation}
For the \(D^0\) weak annihilation, we also reduce the parameter space to
\begin{equation}
\begin{aligned}
&\frac{B_1+\tilde{\Delta}\delta_1+B_2+\tilde{\Delta}\delta_2}{2}=1\pm\frac13,\hspace{1cm}B_1+\tilde{\Delta}\delta_1-B_2-\tilde{\Delta}\delta_2=0\pm0.1,\\
&\frac{\epsilon_1+\tilde{\Delta}\delta_3+\epsilon_2+\tilde{\Delta}\delta_4}{2}=0\pm0.05,\hspace{0.9cm}\epsilon_1+\tilde{\Delta}\delta_3-\epsilon_2-\tilde{\Delta}\delta_4=0\pm0.05,
\end{aligned}
\end{equation}
which is justified by the assumptions that the \(\delta\)s are small and approximate \(SU(3)\) flavor symmetry. Numerically, we obtain for the weak annihilation in \(D_s^+\)
\begin{equation}
 \left(\frac{\overline{\tau}(D_s^+)}{\tau(D^0)}-1\right)_{\text{WA}(D_s^+)}=
 \begin{cases}
   0.12\pm0.06^{(\text{hadronic})}\pm0.02^{\text{(scale)}}\pm0.00^{(\text{parametric})},&\text{pole mass scheme}\\
   0.12\pm0.06^{(\text{hadronic})}\pm0.01^{\text{(scale)}}\pm0.00^{(\text{parametric})},&\overline{\text{MS}}\text{ mass scheme}\\
  \end{cases}
  \label{eq:DspD0WADsp}
\end{equation}
and for the weak annihilation in \(D^0\)
\begin{equation}
 \left(\frac{\overline{\tau}(D_s^+)}{\tau(D^0)}-1\right)_{\text{WA}(D^0)}=
 \begin{cases}
   -0.01\pm0.08^{(\text{hadronic})}\pm0.00^{\text{(scale)}}\pm0.00^{(\text{parametric})},&\text{pole mass scheme}\\
   -0.01\pm0.08^{(\text{hadronic})}\pm0.00^{\text{(scale)}}\pm0.00^{(\text{parametric})},&\overline{\text{MS}}\text{ mass scheme}\\
  \end{cases}.
  \label{eq:DspD0WAD0}
\end{equation}
The contribution (d) from Pauli interference in \(D_s^+\) is Cabibbo-suppressed and should therefore only affect the lifetime difference
at the order of a few percent. It is given by
\begin{equation}
 \begin{aligned}
 \left[\Gamma(D^0)-\Gamma(D_s^+)\right]_\text{PI}&=-\frac{G_F^2m_c^2}{12\pi}f_{D_s}^2M_{D_s}|V_{ud}|^2\Bigg[\vec{F}^{s'u}\cdot\left(\vec{B}^s+\Delta\vec{\delta}\right)\\
 &+\sum\limits_{j=1}^6M_j\left[g_j^{s'u}\left(\rho_j^s+\Delta\delta_{\rho,j}\right)+h_j^{s'u}\left(\sigma_j^s+\Delta\delta_{\sigma,j}\right)\right]\Bigg].
 \end{aligned}
\label{eq:DspPI}
\end{equation}
This yields
\begin{equation}
 \left(\frac{\overline{\tau}(D_s^+)}{\tau(D^0)}-1\right)_\text{(d)}=
 \begin{cases}
   0.04\pm0.05_{\hspace{1.45cm}-0.02}^{(\text{hadronic})\hspace{0.1cm}+0.03\text{ (scale)}}\pm0.00^{(\text{parametric})},&\text{pole mass scheme}\\
   0.06\pm0.05_{\hspace{1.45cm}-0.02}^{(\text{hadronic})\hspace{0.1cm}+0.02\text{ (scale)}}\pm0.00^{(\text{parametric})},&\overline{\text{MS}}\text{ mass scheme}\\
  \end{cases}.
  \label{eq:DspD0c}
\end{equation}
The effect (e) could possibly yield a small contribution because the non-Cabibbo-suppressed Pauli interference is large. We obtain
\begin{equation}
 \begin{aligned}
 &\left[\Gamma(D^0)-\Gamma(D_s^+)\right]_{\text{(e)}}=\frac{G_F^2m_c^2}{12\pi}f_{D}^2M_{D}\Bigg[\vec{F}^{s'u}\cdot\left(\vec{\delta}^{ud}-\frac{f_{D_s}^2M_{D_s}}{f_D^2M_D}\vec{\delta}^{sd}\right)\\
 &+\sum\limits_{j=1}^6M_j\left[g_j^{s'u}\left(\delta_{\rho,j}^{ud}-\frac{f_{D_s}^2M_{D_s}}{f_D^2M_D}\delta_{\rho,j}^{sd}\right)
 +h_j^{s'u}\left(\delta_{\sigma,j}^{ud}-\frac{f_{D_s}^2M_{D_s}}{f_D^2M_D}\delta_{\sigma,j}^{sd}\right)\right]\Bigg].
 \end{aligned}
\label{eq:DspD0e}
\end{equation}
Since nothing is known about the \(\delta\)'s, we can only give a crude estimate about the size of this contribution. If we set \(\delta_1^{ud}=\delta_1^{sd}=0.01\) and
all other \(\delta\)'s to zero, we obtain \(\left[\overline{\tau}(D_s^+)/\tau(D^0)-1\right]_\text{(e)}=0.007\). We do not expect a much larger effect, but at present it can
also not be excluded, and we hence introduce an additional hadronic uncertainty of \(0.05\).
The combination of the various contributions yields
\begin{equation}
 \begin{aligned}
  &\left(\frac{\overline{\tau}(D_s^+)}{\tau(D^0)}\right)_\text{exp}\hspace{0.08cm}=1.289\pm0.019,\\
  &\left(\frac{\overline{\tau}(D_s^+)}{\tau(D^0)}\right)_\text{pole}=1.18\pm0.13^{(\text{hadronic})\hspace{0.15cm}+0.04\hspace{0.1cm}(\text{scale})}_{\hspace{1.5cm}-0.05}\pm0.01^{(\text{parametric})},\\
  &\left(\frac{\overline{\tau}(D_s^+)}{\tau(D^0)}\right)_{\overline{\text{MS}}}\hspace{0.08cm}=1.19\pm0.12^{(\text{hadronic})\hspace{0.15cm}+0.04\hspace{0.1cm}(\text{scale})}_{\hspace{1.5cm}-0.04}\pm0.01^{(\text{parametric})}.
 \end{aligned}
 \label{eq:DspD0HQETresult}
\end{equation}
The theory prediction falls a bit short of the experimental value, but within the theory uncertainty it is well consistent.
%
%
%
%
%
%
%
%
%
\section{Conclusion\label{sec:conclusion}}
In this work, we investigated the validity of the HQE in the charm sector using $D$-meson lifetimes.
In the $B$ sector the HQE is well tested, but doubts about the validity of the \(1/m_c\) expansion have often been voiced.
Because \(m_c\approx m_b/3\), the convergence behavior of the HQE is obviously slower than for $B$ hadrons.
Therefore, we have considered subleading corrections in the \(1/m_c\) expansion as well as NLO QCD corrections.
For \(\tau(D^+)/\tau(D^0)\), we found very good agreement with experimental data:
\begin{equation}
 \begin{aligned}
  &\left(\frac{\tau(D^+)}{\tau(D^0)}\right)_\text{exp}\hspace{2.1cm}=2.536\pm0.019,\\
  &\left(\frac{\tau(D^+)}{\tau(D^0)}\right)_{\overline{\text{MS}}\text{, extr. from \cite{becirevic01}}}\hspace{0.08cm}
  =2.2\pm0.4^{(\text{hadronic})\hspace{0.15cm}+0.3\hspace{0.1cm}(\text{scale})}_{\hspace{1.5cm}-0.7}\pm0.0^{(\text{parametric})}.
 \end{aligned}
\end{equation}
However, an update of hadronic \(\Delta C=0\) matrix elements is direly needed given the advances lattice QCD has made in the past decade.
For the \(\Delta C=2\) matrix elements required in \(D^0-\overline{D}^0\) mixing, such a computation has recently been performed with high accuracy in Ref. \cite{ETMC12}.
We estimate that the remaining scale dependence could be considerably reduced by a NLO calculation of the dimension-7 Wilson coefficients. However, this is
only worthwhile if simultaneous progress on dimension-7 matrix elements is made.\\
The low-momentum release decay $D_s^+\ra\eta'(958)\rho^+$ poses a potential threat to the HQE description of the $D_s^+$ lifetime. Thus, we would expect a possible 
failure of the HQE to be most apparent here. We found that the HQE result for the \(D_s^+-D^0\) lifetime ratio falls slightly
short of the experimental value, but it is consistent within hadronic uncertainties:
\begin{equation}
 \begin{aligned}
  &\left(\frac{\overline{\tau}(D_s^+)}{\tau(D^0)}\right)_\text{exp}\hspace{0.08cm}=1.289\pm0.019,\\
  &\left(\frac{\overline{\tau}(D_s^+)}{\tau(D^0)}\right)_{\overline{\text{MS}}}\hspace{0.08cm}=1.19\pm0.12^{(\text{hadronic})\hspace{0.15cm}+0.04\hspace{0.1cm}(\text{scale})}_{\hspace{1.5cm}-0.04}\pm0.01^{(\text{parametric})}.
 \end{aligned}
\end{equation}
Presently, however, this does not exclude possible large violations of the HQE. A nonperturbative determination of the hadronic matrix elements could provide a more stringent upper bound.
In addition, this would offer the unique possibility to use semileptonic decays, where the momentum release is large, to extract information on the nonvalence contractions.
This is not possible in $B$ decays, because the semileptonic weak annihilation is doubly CKM suppressed and the difference of the semileptonic widths is too small to be measured experimentally.\\
The subleading corrections to the lifetimes are large \(\left(\approx30\%\text{ QCD, }\approx50\%\text{ }1/m_c\right)\), but still allow a description within the realm of perturbation theory.
Similar behavior was found in an earlier study of \(D^0-\overline{D}^0\) mixing \cite{bobrowski11}. The analysis of the \(\mu_1\) dependence of our results suggests that perturbation theory
breaks down below about \(1\text{ GeV}\) but still works at the charm scale. In combination with the intriguing agreement of the standard-model HQE prediction for \(\Delta\Gamma_s\) with
experiment in spite of the small momentum release, this justifies confidence in the validity of the HQE. Still, lattice inputs are crucial to confirming this view.

\begin{acknowledgments}
We are grateful to U. Nierste for giving us access to a program from the computation in Ref. \cite{beneke02} that allowed us to check the results of Appendix \ref{sec:SLWA},
for clarifying discussions, and for carefully reading the manuscript. We are thankful to V. Khoze, M. Voloshin, I. Bigi and in particular N. Uraltsev for various comments
on the history of $D$-meson lifetimes.
T.R. would like to thank G. Kirilin, J. Rohrwild and F. Krinner for helpful discussions, the IPPP Durham and KIT Karlsruhe for hospitality, and M. Beneke and A. Ibarra for support during his thesis.
\end{acknowledgments}

\appendix
\section{NLO Wilson coefficients for the semileptonic weak annihilation\label{sec:SLWA}}
We decompose the Wilson coefficients as
\begin{equation}
 F^{qq'}=F^{qq',(0)}+\frac{\alpha_s}{4\pi}F^{qq',(1)},\dots
\label{eq:LONLOdecomp}
\end{equation}
The Wilson coefficients for the semileptonic weak annihilation in the scheme of Ref. \cite{ROME} read
\begin{equation}
\begin{aligned}
 &\tilde{F}^{\nu\mu,(0)}(z)=-(1-z)^2\left(1+\frac{z}{2}\right),\hspace{1cm}\tilde{F}_S^{\nu\mu,(0)}(z)=(1-z)^2\left(1+2z\right),\\
 &\tilde{G}^{\nu\mu,(0)}(z)=0,\hspace{3.95cm}\tilde{G}_S^{\nu\mu,(0)}(z)=0,
\end{aligned}
 \label{eq:dim6slWALOcoefficients}
\end{equation}
and
\begin{equation}
 \begin{aligned}
  &\tilde{F}^{\nu\mu,(1)}(z)=\frac{4}{3} (1-z)^2 (2+z) \left[4+3 \log\left(\frac{\mu_0}{m_c}\right)\right],\\
  &\tilde{F}_S^{\nu\mu,(1)}(z)=-\frac{8}{3} (1-z)^2 \left[4+5 z+3(1+2 z)\log\left(\frac{\mu_0}{m_c}\right)\right],\\
  &\tilde{G}^{\nu\mu,(1)}(z)=\frac{1}{18}\left[(1-z) (-205-7z+110 z^2)-6 z^2 (6+11 z) \log(z)\right]\\
  &\hspace{2cm}-3(1-z)^2(2+z)\left[\log\left(\frac{\mu_0}{m_c}\right)-\log(1-z)\right],\\
  &\tilde{G}_S^{\nu\mu,(1)}(z)=\frac{1}{9}(1-z)(95+104z-211z^2)-\frac{4}{3} z^2 (12-11 z)\log(z)\\
  &\hspace{2cm}+6 (1-z)^2 (1+2 z)\left[\log\left(\frac{\mu_0}{m_c}\right)-\log(1-z)\right],
 \end{aligned}
 \label{eq:slWAcoeffs}
\end{equation}
where \(z=m_\mu^2/m_c^2\). Note that our convention for the Wilson coefficients differs from that of Ref.  \cite{ROME} by a factor of 3.
The details of the calculation have been described in Refs. \cite{beneke02,ROME}.
We have checked the correctness of these coefficients against intermediate results from the computation in Ref. \cite{beneke02} kindly
made available to us by Ulrich Nierste.
\section{Operator basis and Wilson coefficients for dimension seven\label{sec:dim7}}
We use the following basis for the dimension-7 operators
\begin{equation}
 \begin{aligned}
  &P_1^q=\frac{m_q}{m_c}\bar{c}(1-\gamma_5)q\otimes\bar{q}(1-\gamma_5)c,\hspace{1cm}P_2^q=\frac{m_q}{m_c}\bar{c}(1+\gamma_5)q\otimes\bar{q}(1+\gamma_5)c,\\
  &P_3^q=\frac{1}{m_c^2}\bar{c}\overleftarrow{D}_\rho\gamma_\mu(1-\gamma_5)D^\rho q\otimes\bar{q}\gamma^\mu(1-\gamma_5)c,\\
  &P_4^q=\frac{1}{m_c^2}\bar{c}\overleftarrow{D}_\rho(1-\gamma_5)D^\rho q\otimes\bar{q}(1+\gamma_5)c,\\
  &P_5^q=\frac{1}{m_c}\bar{c}\gamma_\mu(1-\gamma_5)q\ \bar{q}\gamma^\mu(1-\gamma_5)\left(i\slashed{D}\right)c,\\
  &P_6^q=\frac{1}{m_c}\bar{c}(1-\gamma_5)q\ \bar{q}(1+\gamma_5)\left(i\slashed{D}\right)c.
 \end{aligned}
\label{eq:dim7ops}
\end{equation}
The \(S_i^q\) are the corresponding color octet operators obtained by inserting \(T^A\) in the two currents of the respective color singlet operators.
As discussed at the end of Sec. \ref{sec:HQE}, the operators $P_{5,6}^q$ and $S_{5,6}^q$ only occur in HQET and are absent if we use QCD operators.
We decompose the Wilson coefficients for dimension seven defined in Eq. \eqref{eq:T4s} as
\begin{equation}
 g_i^{qq'}=C_1^2 g_{i,11}^{qq',(0)}+C_1C_2g_{i,12}^{qq',(0)}+C_2^2g_{i,22}^{qq',(0)}+\mathcal{O}(\alpha_s).
\end{equation}
As results for the LO coefficients we obtain
\begin{equation}
\begin{aligned}
 &\frac13g_{1,11}^{sd,(0)}(z)=\frac12g_{1,12}^{sd,(0)}(z)=3g_{1,22}^{sd,(0)}(z)=\frac12h_{1,22}^{sd,(0)}(z)=-(1-z)^2(1+2z),\\
 &\frac13g_{2,11}^{sd,(0)}(z)=\frac12g_{2,12}^{sd,(0)}(z)=3g_{2,22}^{sd,(0)}(z)=\frac12h_{2,22}^{sd,(0)}(z)=-(1-z)^2(1+2z),\\
 &\frac13g_{3,11}^{sd,(0)}(z)=\frac12g_{3,12}^{sd,(0)}(z)=3g_{3,22}^{sd,(0)}(z)=\frac12h_{3,22}^{sd,(0)}(z)=2(1-z)\left(1+z+z^2\right),\\
 &\frac13g_{4,11}^{sd,(0)}(z)=\frac12g_{4,12}^{sd,(0)}(z)=3g_{4,22}^{sd,(0)}(z)=\frac12h_{4,22}^{sd,(0)}(z)=-12z^2(1-z),\\
 &\frac13g_{1,11}^{ss,(0)}(z)=\frac12g_{1,12}^{ss,(0)}(z)=3g_{1,22}^{ss,(0)}(z)=\frac12h_{1,22}^{ss,(0)}(z)=-\sqrt{1-4z}(1+2z),\\
 &\frac13g_{2,11}^{ss,(0)}(z)=\frac12g_{2,12}^{ss,(0)}(z)=3g_{2,22}^{ss,(0)}(z)=\frac12h_{2,22}^{ss,(0)}(z)=-\sqrt{1-4z}(1+2z),\\
 &\frac13g_{3,11}^{ss,(0)}(z)=\frac12g_{3,12}^{ss,(0)}(z)=3g_{3,22}^{ss,(0)}(z)=\frac12h_{3,22}^{ss,(0)}(z)=\frac{2}{\sqrt{1-4z}}\left[1-2z(1+z)\right],\\
 &\frac13g_{4,11}^{ss,(0)}(z)=\frac12g_{4,12}^{ss,(0)}(z)=3g_{4,22}^{ss,(0)}(z)=\frac12h_{4,22}^{ss,(0)}(z)=-\frac{24z^2}{\sqrt{1-4z}},\\
 &6g_{3,11}^{su,(0)}(z)=g_{3,12}^{su,(0)}(z)=6g_{3,22}^{su,(0)}(z)=h_{3,11}^{su,(0)}(z)=h_{3,22}^{su,(0)}(z)=12(1-z)(1+z),\\
 &h_{i,11}^{sd,(0)}=h_{i,12}^{sd,(0)}=h_{i,11}^{ss,(0)}=h_{i,12}^{ss,(0)}=0,\\
 &g_{1,ij}^{su,(0)}=g_{2,ij}^{su,(0)}=g_{4,ij}^{su,(0)}=h_{1,ij}^{su,(0)}=h_{2,ij}^{su,(0)}=h_{4,ij}^{su,(0)}=h_{3,12}^{su,(0)}=0.\\
\end{aligned}
 \label{eq:dim7LOcoefficients}
\end{equation}
The coefficients \(g_{i,jk}^{ds,(0)}\) are identical to \(g_{i,jk}^{sd,(0)}\) because of the symmetry under interchange of the masses in the loops,
and \(g_{i,jk}^{dd,(0)}\) and \(g_{i,jk}^{du,(0)}\) follow by setting \(z=0\) in \(g_{i,jk}^{sd,(0)}\) and \(g_{i,jk}^{su,(0)}\) respectively.
As with the dimension-6 operators, the coefficients of the weak annihilation in \(D_{(s)}^+\) with quarks in the loop are identical to those of
the weak annihilation in \(D^0\), when we interchange \(C_1\) and \(C_2\). For the semileptonic weak annihilation, we obtain
\begin{equation}
\begin{aligned}
 &\tilde{g}_1^{\nu\mu,(0)}(z)=-(1-z)^2(1+2z),\hspace{1.55cm}\tilde{g}_2^{\nu\mu,(0)}(z)=-(1-z)^2(1+2z),\\
 &\tilde{g}_3^{\nu\mu,(0)}(z)=2(1-z)\left(1+z+z^2\right),\hspace{1cm}\tilde{g}_4^{\nu\mu,(0)}(z)=-12z^2(1-z),\\
 &\tilde{h}_i^{\nu\mu,(0)}(z)=0,
\end{aligned}
 \label{eq:dim7slWALOcoefficients}
\end{equation}
where \(z=m_\mu^2/m_c^2\) and the \(\tilde{g}_i^{\nu e,(0)}\) are given by setting \(z=0\). The additional coefficients that arise when we use
HQET operators are given by the relation
\begin{equation}
\begin{aligned}
&\left(g_{5,ij}^{sq,(0)}\right)_\text{HQET}=\left(F_{ij}^{sq,(0)}\right)_\text{QCD},\hspace{1cm}
\left(g_{6,ij}^{sq,(0)}\right)_\text{HQET}=\left(F_{S,ij}^{sq,(0)}\right)_\text{QCD},\\
&\left(h_{5,ij}^{sq,(0)}\right)_\text{HQET}=\left(G_{ij}^{sq,(0)}\right)_\text{QCD},\hspace{1cm}
\left(h_{6,ij}^{sq,(0)}\right)_\text{HQET}=\left(G_{S,ij}^{sq,(0)}\right)_\text{QCD}.\\
\end{aligned}
 \label{eq:dim7LOcoefficientsHQET}
\end{equation}

\section{Parametrization of the matrix elements\label{sec:parametrization}}
We parametrize the matrix elements of the dimension-6 QCD operators following Ref. \cite{beneke02}:
\begin{equation}
 \begin{aligned}
\frac{\Braket{D^+|Q^{d}-Q^u|D^+}}{2M_{D}}=\frac{f_{D}^2M_{D}}{2}B_1,\hspace{1cm} &\frac{\Braket{D^+|Q_S^{d}-Q_S^u|D^+}}{2M_{D}}=\frac{f_{D}^2M_{D}}{2}B_2,\\
\frac{\Braket{D^+|T^{d}-T^u|D^+}}{2M_{D}}=\frac{f_{D}^2M_{D}}{2}\epsilon_1,\hspace{1cm} &\frac{\Braket{D^+|T_S^{d}-T_S^u|D^+}}{2M_{D}}=\frac{f_{D}^2M_{D}}{2}\epsilon_2.
\end{aligned}
\label{eq:QCDparametrization}
\end{equation}
For \(q=u,d\), the dimension-7 operators \(P_1^q,P_2^q,S_1^q,S_2^q\) vanish identically if we neglect the up and down quark masses.
The operators \(P_5^q,P_6^q,S_5^q,S_6^q\) do not arise in QCD. We choose the following parametrization for the remaining ones:
\begin{equation}
 \begin{aligned}
&\frac{\Braket{D^+|P_3^{d}-P_3^u|D^+}}{2M_{D}}=-\frac{f_{D}^2M_{D}}{2}\frac{M_D-m_c}{m_c}\rho_3+\mathcal{O}\left(1/m_c^2\right),\\
&\frac{\Braket{D^+|P_4^{d}-P_4^u|D^+}}{2M_{D}}=\frac{f_{D}^2M_{D}}{2}\frac{M_D-m_c}{m_c}\rho_4+\mathcal{O}\left(1/m_c^2\right),\\
&\frac{\Braket{D^+|S_3^{d}-S_3^u|D^+}}{2M_{D}}=-\frac{f_{D}^2M_{D}}{2}\frac{M_D-m_c}{m_c}\sigma_3+\mathcal{O}\left(1/m_c^2\right),\\
&\frac{\Braket{D^+|S_4^{d}-S_4^u|D^+}}{2M_{D}}=\frac{f_{D}^2M_{D}}{2}\frac{M_D-m_c}{m_c}\sigma_4+\mathcal{O}\left(1/m_c^2\right).
\end{aligned}
\label{eq:QCDparametrization7}
\end{equation}
This parametrization is inspired by the \textit{vacuum saturation approximation} (VSA), where \(B_1=1\), \(B_2=1+2\left(M_D-m_c\middle)\right/m_c+\mathcal{O}\left(1/m_c^2\right)\), and
\(\epsilon_1=\epsilon_2=0\) for dimension six, and \(\rho_3=\rho_4=1\) and \(\sigma_3=\sigma_4=0\) for dimension seven.\\
For HQET operators, we use a parametrization following Ref. \cite{ROME}, where the contributions of valance and nonvalence operators
are distinguished explicitly. We parametrize the matrix elements of the nonvalence operators \((q\neq q')\) by
\begin{equation}
\begin{aligned}
\frac{\Braket{D_q|Q^{q'}|D_q}}{2M_{D_q}}=\frac{f_{D_q}^2M_{D_q}}{2}\delta_1^{qq'},\hspace{1cm} &\frac{\Braket{D_q|Q_S^{q'}|D_q}}{2M_{D_q}}=\frac{f_{D_q}^2M_{D_q}}{2}\delta_2^{qq'},\\
\frac{\Braket{D_q|T^{q'}|D_q}}{2M_{D_q}}=\frac{f_{D_q}^2M_{D_q}}{2}\delta_3^{qq'},\hspace{1cm} &\frac{\Braket{D_q|T_S^{q'}|D_q}}{2M_{D_q}}=\frac{f_{D_q}^2M_{D_q}}{2}\delta_4^{qq'},
\end{aligned}
\label{eq:nonvalenceoperators}
\end{equation}
and the matrix elements of the valence operators \((q=q')\) by
\begin{equation}
\begin{aligned}
\frac{\Braket{D_q|Q^{q}|D_q}}{2M_{D_q}}=\frac{f_{D_q}^2M_{D_q}}{2}\left(B_1^q+\delta_1^{qq}\right),\hspace{1cm} &\frac{\Braket{D_q|Q_S^{q}|D_q}}{2M_{D_q}}=\frac{f_{D_q}^2M_{D_q}}{2}\left(B_2^q+\delta_2^{qq}\right),\\
\frac{\Braket{D_q|T^{q}|D_q}}{2M_{D_q}}=\frac{f_{D_q}^2M_{D_q}}{2}\left(\epsilon_1^q+\delta_3^{qq}\right),\hspace{1cm} &\frac{\Braket{D_q|T_S^{q}|D_q}}{2M_{D_q}}=\frac{f_{D_q}^2M_{D_q}}{2}\left(\epsilon_2^q+\delta_4^{qq}\right).
\end{aligned}
\label{eq:valenceoperators}
\end{equation}
In the VSA, we find \(B_1^q=B_2^q=1\), while the \(\epsilon_1^q\), \(\epsilon_2^q\), and all the \(\delta\)'s vanish. 
We proceed in the same way for the dimension-7 matrix elements of nonvalence operators
\begin{equation}
\begin{aligned}
\frac{\Braket{D_q|P_i^{q'}|D_q}}{2M_{D_q}}=-\frac{f_{D_q}^2M_{D_q}}{2}\frac{m_{q'}}{m_c}\delta_{\rho,i}^{qq'},\hspace{4.1cm}i=1,2\\
\frac{\Braket{D_q|P_i^{q'}|D_q}}{2M_{D_q}}=(-1)^i\frac{f_{D_q}^2M_{D_q}}{2}\frac{M_{D_q}-m_c}{m_c}\delta_{\rho,i}^{qq'},\hspace{2.5cm}i=3,4\\
\frac{\Braket{D_q|P_i^{q'}|D_q}}{2M_{D_q}}=(-1)^i\frac{f_{D_q}^2M_{D_q}}{2}\frac{M_{D_q}-m_c-m_{q'}}{m_c}\delta_{\rho,i}^{qq'},\hspace{1.4cm}i=5,6\\
\end{aligned}
\label{eq:nonvalenceoperatorsdim7}
\end{equation}
and valence operators
\begin{equation}
\begin{aligned}
\frac{\Braket{D_q|P_i^{q}|D_q}}{2M_{D_q}}=-\frac{f_{D_q}^2M_{D_q}}{2}\frac{m_{q}}{m_c}\left(\rho_i^q+\delta_{\rho,i}^{qq}\right),\hspace{4.05cm}i=1,2\\
\frac{\Braket{D_q|P_i^{q}|D_q}}{2M_{D_q}}=(-1)^i\frac{f_{D_q}^2M_{D_q}}{2}\frac{M_{D_q}-m_c}{m_c}\left(\rho_i^q+\delta_{\rho,i}^{qq}\right),\hspace{2.35cm}i=3,4\\
\frac{\Braket{D_q|P_i^{q}|D_q}}{2M_{D_q}}=(-1)^i\frac{f_{D_q}^2M_{D_q}}{2}\frac{M_{D_q}-m_c-m_{q}}{m_c}\left(\rho_i^q+\delta_{\rho,i}^{qq}\right),\hspace{1.4cm}i=5,6\\
\end{aligned}
\label{eq:valenceoperatorsdim7}
\end{equation}
The color octet operators are parametrized by Eqs. \eqref{eq:nonvalenceoperatorsdim7} and \eqref{eq:valenceoperatorsdim7} with the replacements \(P\ra S\) and
\(\rho\ra\sigma\). This is chosen such that in the vacuum insertion \(\rho_i^q=1\), and all the \(\sigma\)'s and \(\delta\)'s vanish.
\newpage
\section{Inputs for the numerical evaluation}
\label{sec:inputs}
\begin{table}[htbc]
\centering
 \begin{tabular}{|c|c|c||c|c|c|}\hline
\(\overline{m}_c(\overline{m}_c)\)&\((1.275\pm0.025)\text{ GeV}\)&\cite{pdg}		&\(M_{D^0}\)&\((1864.86\pm0.13)\text{ MeV}\)&\cite{pdg}\\ \hline
\(\overline{m}_s(2\text{ GeV})\)&\((95\pm5)\text{ MeV}\)					&\cite{pdg}&\(M_{D^+}\)&\((1869.62\pm0.15)\text{ MeV}\)&\cite{pdg}\\ \hline
\(\overline{m}_b(\overline{m}_b)\)&\((4.18\pm0.03)\text{ GeV}\)&\cite{pdg}		&\(M_{D_s^+}\)&\((1968.49\pm0.32)\text{ MeV}\)&\cite{pdg}\\ \hline
\(\alpha_s(M_Z)\)&\((0.1184\pm0.0007)\)&\cite{pdg}			&\(\tau(D^0)\)&\((410.1\pm1.5)\times10^{-15}\text{ s}\)&\cite{pdg}\\ \hline
\(f_D\)&\((212.7\pm3.2)\text{ MeV}\)&\cite{rong12}					&\(\tau(D^+)\)&\((1040\pm7)\times10^{-15}\text{ s}\)&\cite{pdg}\\ \hline
\(f_{D_s}\)&\((260.0\pm5.4)\text{ MeV}\)&\cite{pdg}					&\(\tau(D_s^+)\)&\((500\pm7)\times10^{-15}\text{ s}\)&\cite{pdg}\\ \hline
\(|V_{us}|\)&\(0.2252\pm0.0009\)&\cite{pdg}						&\(\text{Br}(D^0\ra Xe^+\nu)\)&\((6.49\pm0.11)\%\)&\cite{pdg}\\ \hline
\(|V_{ub}|\)&\((4.15\pm0.49)\times10^{-3}\)&\cite{pdg}					&\(\text{Br}(D^+\ra Xe^+\nu)\)&\((16.07\pm0.30)\%\)&\cite{pdg}\\ \hline
\(|V_{cb}|\)&\((40.9\pm1.1)\times10^{-3}\)&\cite{pdg}					&\(\text{Br}(D_s^+\ra Xe^+\nu)\)&\((6.5\pm0.4)\%\)&\cite{pdg}\\ \hline
\(\delta_\text{CKM}\)&\((68_{-11}^{+10})^\circ\)&\cite{pdg}				&\(\text{Br}(D_s^+\ra \tau^+\nu)\)&\((5.43\pm0.31)\%\)&\cite{pdg}\\ \hline
\(m_\mu\)&\(105.658\text{ MeV}\)&\cite{pdg}&&&\\ \hline
 \end{tabular}
 \caption{Input parameters for the numerical evaluation\label{tab:expinputs}}
\end{table}


\end{document}